\documentclass[12pt]{article}
\usepackage[utf8x]{inputenc}
\usepackage[english]{babel}
\usepackage{natbib}
\usepackage{graphicx}
\usepackage{amsmath}
\usepackage{amsfonts}
\usepackage{algorithm}
\usepackage{algorithmic}
\usepackage{hyperref}
\usepackage{authblk}
\newcommand{\Real}{\mathbb{R}}
\newcommand{\T}{^\top}

\title{Weighted likelihood estimation of multivariate location and scatter}
\author[1]{Claudio Agostinelli}
\author[2]{Luca Greco}

\affil[1]{Department of Mathematics,University of Trento, Italy, \texttt{claudio.agostinelli@unitn.it}}
\affil[2]{DEMM Department, University of Sannio, Italy \texttt{luca.greco@unisannio.it}}

\begin{document}

\date{\today}
\maketitle

\begin{abstract}
A novel approach to obtain weighted likelihood estimates of multivariate location and scatter is discussed. 
A weighting scheme is proposed that is based on the distribution of the Mahalanobis distances rather than the distribution of the data at the assumed model. 
This strategy allows to avoid the curse of dimensionality affecting non-parametric density estimation, that is involved in the construction of the weights through the Pearson residuals \citep{markatou1998}. 
Then, weighted likelihood based outlier detection rules and robust dimensionality reduction techniques are developed.
The effectiveness of the methodology is illustrated through some numerical studies and real data examples. 

\noindent \textbf{Keywords}: Dimensionality Reduction; Discriminant Analysis; Mahalanobis distance; Multivariate Normal; Outlier detection; Pearson residuals; PCA; Robustness; Weighted Likelihood \\
\noindent \textbf{Mathematics Subject Classification (2000)}: MSC 62F35; MSC 62G35; MSC 62H25; MSC 62H30 
\end{abstract}

\newpage

\section{Introduction}
Several multivariate techniques are based on the the assumption of multivariate normality and the use of the sample mean vector and covariance matrix. It is well known that small departures from model assumptions may invalidate classical analysis completely \citep{maronna2006robust, huber2robust}.
Such departures result in data inadequacies that are typically observed in the form of several outliers.
Outliers can be defined as observations that are highly unlikely to occur under the assumed model \citep {markatou1998}. In other words, outliers {\it contaminate} the data with respect to (w.r.t.) the postulated model. Contamination in the data may have dramatic effects on all those techniques based on multivariate estimation of location and scatter, such as Principal Component Analysis or Discriminant Analysis, for instance.
On the contrary, by supplying robust estimates of multivariate location and covariance, one could rely on multivariate techniques that are resistant to contamination \citep{hubert2008high}. Furthermore, the appropriate use of robust estimators may also lead to detect outliers, find unexpected structures in the data and explore the types of occurred departures.
There is a growing literature on robust multivariate estimation. The reader is pointed to 
the book by \cite{farcomeni2016robust} for a recent account on multivariate settings.

Robust estimates  of multivariate location and covariance are obtained by attaching a weight to each data point
in order to bound the effect of possible outliers on the resulting fit. Weights are determined according to an outlyingness measure, that is a measure of the distance of the multivariate data point from the robust fit.
In summary, we can consider three main classes of estimators.
\begin{enumerate}
\item Estimators based on hard trimming: weights are 0-1 and outliers are trimmed. The final estimate is based on a subset of the original data points, whose size is tuned by the user. The Minimum Covariance Determinant (MCD) is undoubtedly one of the most popular techniques \citep{rousseeuw1985, croux1999}.
\item Estimators based on adaptive hard trimming: outliers are trimmed but the final sample is determined adaptively by the data. The main tool is represented by the Forward Search (FS, see \cite{riani2009, atkinson2012} for a recent account).
\item Estimators based on soft trimming: outliers are down-weighted, with weights varying in $[0,1]$, and the final estimate consists of a weighted mean and weighted covariance matrix. This feature characterizes M-estimators and related methods such as S-estimators \citep{lopuhaa1989relation} and MM-estimators \citep{salibian2006}, but also the weighted likelihood estimator (WLE, \cite{markatou1998}). In this class we also include those methods stemming from projection of multivariate data onto univariate directions, as the Stahel-Donoho estimator, for instance. 
\end{enumerate}  

The weighting strategy that characterizes the MCD, the FS and M-estimation is based on the inspection of the Mahalanobis distances. 
Let $y=(y_1,y_2,\ldots,y_p)^{\T}$ denote a $p$-variate observation, $p>1$, sampled from a multivariate normal model, $Y\sim N_p(\mu,\Sigma)$ with mean vector $\mu=(\mu_1,\mu_2,\ldots,\mu_p)^{\T}$ and $p\times p$ covariance matrix $\Sigma$. The Mahalanobis distance is  
\begin{equation}
\label{dist}
d(y;\mu,\Sigma)=\sqrt{(y-\mu)^{\T}\Sigma^{-1}(y-\mu)} \ .
\end{equation}
Let $(\hat\mu,\hat\Sigma)$ be a robust estimate of location and scatter, then data points are discarded or down-weighted according to their distance $d(y;\hat\mu,\hat\Sigma)$ from the robust fit: the larger the robust distance the closer to zero the weight and more likely the point will be treated as an outlier.

In a different fashion, the computation of the WLE is not based on such robust distances, but outlyingness is measured according to the agreement between the data, summarized by a non parametric density estimate,
 and the assumed multivariate normal model.
Actually, the weighting scheme based on the computation of a multivariate density estimate becomes troublesome for large dimensions, because of the curse of dimensionality  \citep{huber1985projection, scott1991}. With growing dimensions the data are more sparse and kernel density estimation may become unfeasible. 
To the best of our knowledge, non parametric kernel estimation is implemented in a statistical software as R, up to three dimensions. The method by \cite{duong2007ks} is an exception since it allows to get a non parametric kernel estimate up to six dimensions. The reader is pointed to \cite{deng2011} for a comparison of several density estimation methods available from R.

This feature represents a serious limitation of the weighted likelihood methodology in a multivariate framework. Such a restriction is much more annoying since all the other multivariate estimators that we have mentioned so far are well behaved in large dimensions, at least up to $p=100$. It is worth to stress here that we only consider the case where the sample size $n$ is larger than the dimension $p$.

In this paper, a novel approach to overcome this hindrance is presented. We introduce a weighting algorithm that is still based on non parametric density estimation, but now it is driven from (robust) distances rather than from the data. Hence, the new algorithm handles a univariate kernel density estimate.
We obtain multivariate estimates of location and covariance that are consistent and fully efficient at the assumed multivariate normal model, robust w.r.t. the presence of outliers, with weights that depend on robust distances as well as for the MCD, the FS and M-type estimators.

The rest of the paper is structured as follows. Some background on the weighted likelihood is given in Section 2. The new weighting algorithm is introduced in Section 3, whereas an outlier detection rule is given in Section 4. Some numerical studies are given in Section 5 and real data examples concerning estimation, outlier detection, principal component analysis and discriminant analysi are discussed in Section 6.   

\section{Background}
Let $y = (y_1, \cdots, y_n)$ be a random sample from a random variable $Y$ with unknown probability (density) function $m(y; \theta)$,  $\theta \in \Theta \subset \mathbb{R}^p$, with $p\geq 1$ and let $\hat{M}_n$ be the empirical distribution function. A weighted likelihood estimate is defined as the root of the weighted likelihood estimating equation 
\begin{equation*} 
\label{wlee}
\sum_{i=1}^n w_i \ s(y_i; \theta) = 0 \ ,
\end{equation*}
where $s(y_i; \theta)$ denotes the $i$-th contribution to the score function and
the weight $w_i$ is defined as 
\begin{equation}
\label{weight}
w=w(y; \theta, \hat{M}_n) = \frac{\left[A(\delta(y; \theta, \hat{M}_n)) + 1\right]^+}{\delta(y; \theta, \hat{M}_n) + 1} \ ,
\end{equation}
where $[\cdot]^+$ denotes the positive part.
The function $\delta(y; \theta, \hat{M}_n)$ is the Pearson residual function \citep {markatou1998}
\begin{equation*}
\label{residual}
\delta(y; \theta, \hat{M}_n) = \frac{\hat m_n(y) }{m^*(y; \theta)}-1
\end{equation*}
and 
 $A(\cdot)$ is the Residual Adjustment Function (RAF, \cite{lindsay1994,park+basu+lindsay+2002}).
The Pearson residuals are evaluated by comparing 
a non parametric density estimate 
$$\hat m_n(y)=\frac{1}{n} \sum_{i=1}^nk(y; y_i, h)$$
based on the kernel $k(\cdot;\cdot)$ with bandwidth $h$, and a smoothed version of the model density 
$$m^*(y; \theta)=\int_\mathcal{Y} k(y; t, h) m(t;\theta) dt$$ based on the same kernel function.  Model smoothing leads to the desired asymptotic behavior of the weights, that will be described below, but, in finite samples, for large sample sizes relative to the dimensionality of the parameter space, its presence/absence does not have an important impact on the estimation process.

The RAF plays the role to bound the effect of large residuals on the fitting procedure, as well as the Huber and Tukey-bisquare function bound large distances in M-type estimation.
Here, we consider the families of RAF based on the Power Divergence Measure
\begin{equation*}
A_\tau(\delta) = \left\{
\begin{array}{lc}
\tau \left( (\delta + 1)^{1/\tau} - 1 \right) & \tau < \infty \\
\log(\delta + 1) & \tau \rightarrow \infty 
\end{array}
\right .
\end{equation*}
Special cases are maximum likelihood ($\tau = 1$, as the weights become all equal to one), Hellinger distance ($\tau = 2$), Kullback--Leibler divergence ($\tau \rightarrow \infty$) and Neyman's Chi--Square ($\tau=-1$). 
An alternative is represented by the families of RAF based on the Generalized Kullback-Leibler divergence (see \cite{cressie+1984, cressie+1988, park+basu+2003} and references therein).

When the model is correctly specified, the Pearson residual function evaluated at the true parameter value converges almost surely to zero, whereas, otherwise, for each value of the parameters, large Pearson residuals detect regions where the observation is unlikely to occur under the assumed model. Hence, those observations lying in such regions are attached a weight that decreases with increasing Pearson residual.  Large Pearson residuals and small weights will correspond to data points that are likely to be outliers.

Under classical regularity assumptions regarding the model, the kernel, the RAF and the weight function and a correctly specified model, 
\begin{enumerate}
\item the WLE $\hat\theta$ is consistent and first order efficient \citep{markatou1998}
\item the robustness weights satisfy $$\sup_y|w(y, \hat\theta, \hat M_n)-1|\stackrel{p}{\rightarrow}0$$  \citep{agostinelli+2002, agostinelli2013} 
\item the weighted likelihood ratio test and its asymptotically equivalent versions share the same first order asymptotic properties of their genuine likelihood counterparts \citep{agostinelli2001test} . 
\end{enumerate}

\section{Weighted likelihood estimation based on robust distances}
At the multivariate normal model, the Mahalanobis distance given in (\ref{dist}) satisfies 
$$d^2(Y,\mu,\Sigma)\sim\chi^2_p$$
at the true parameter values. In order to define a set of weights whose computation does not need the evaluation of a multivariate kernel density estimate and does not suffer from any problem due to large dimensionality, 
we suggest to focus on the distribution of squared distances rather than observations.
In other words, the weighting scheme will be based on Person residuals aiming at measuring the degree of agreement between a univariate kernel density estimate based on the vector of squared distances and their underlying $\chi^2_p$ distribution at the assumed multivariate normal model. Then, this strategy leads to down-weight those observations that exhibit a large distance from the robust fit.

\begin{figure}[t]
\centering
\includegraphics[width=0.32\textwidth]{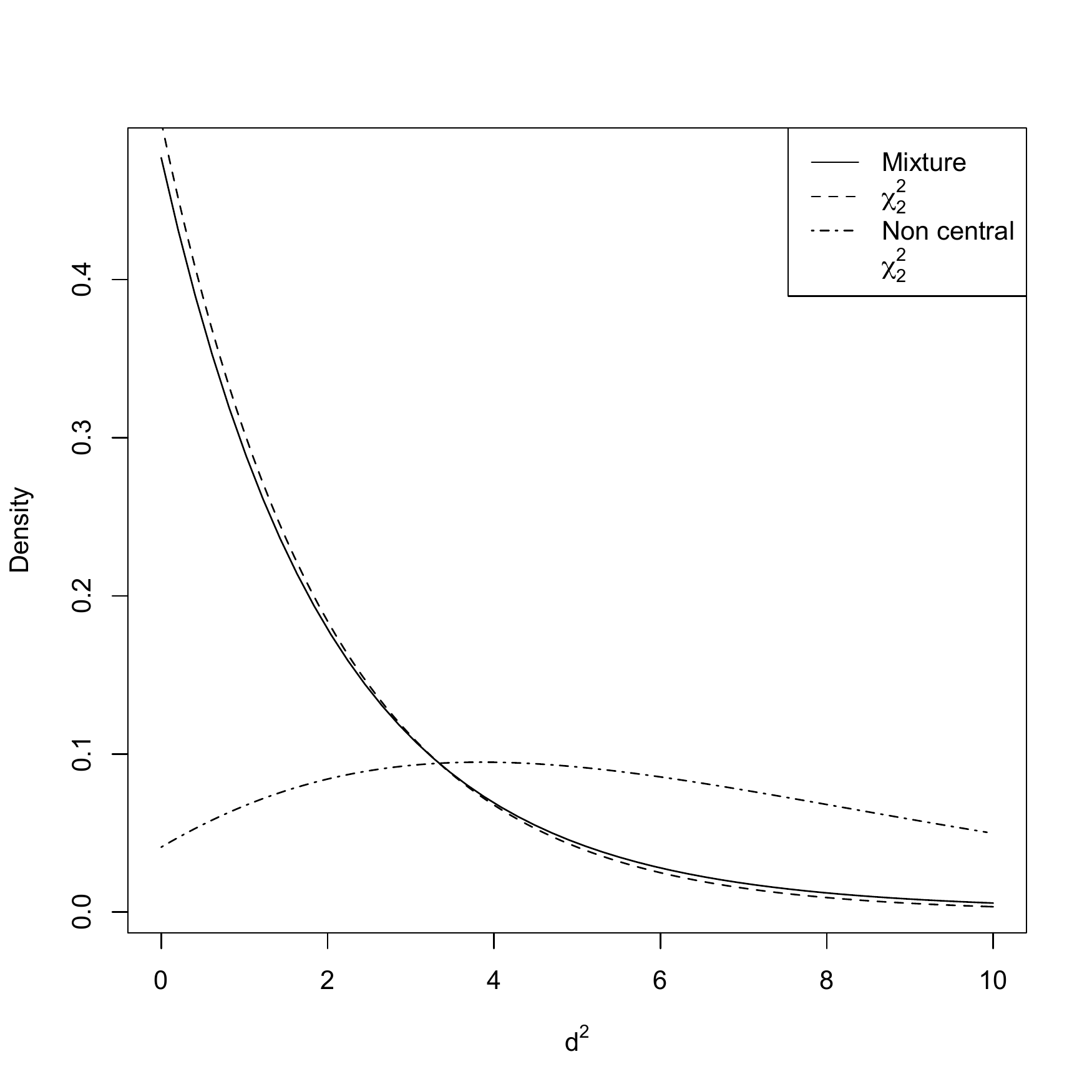}
\includegraphics[width=0.32\textwidth]{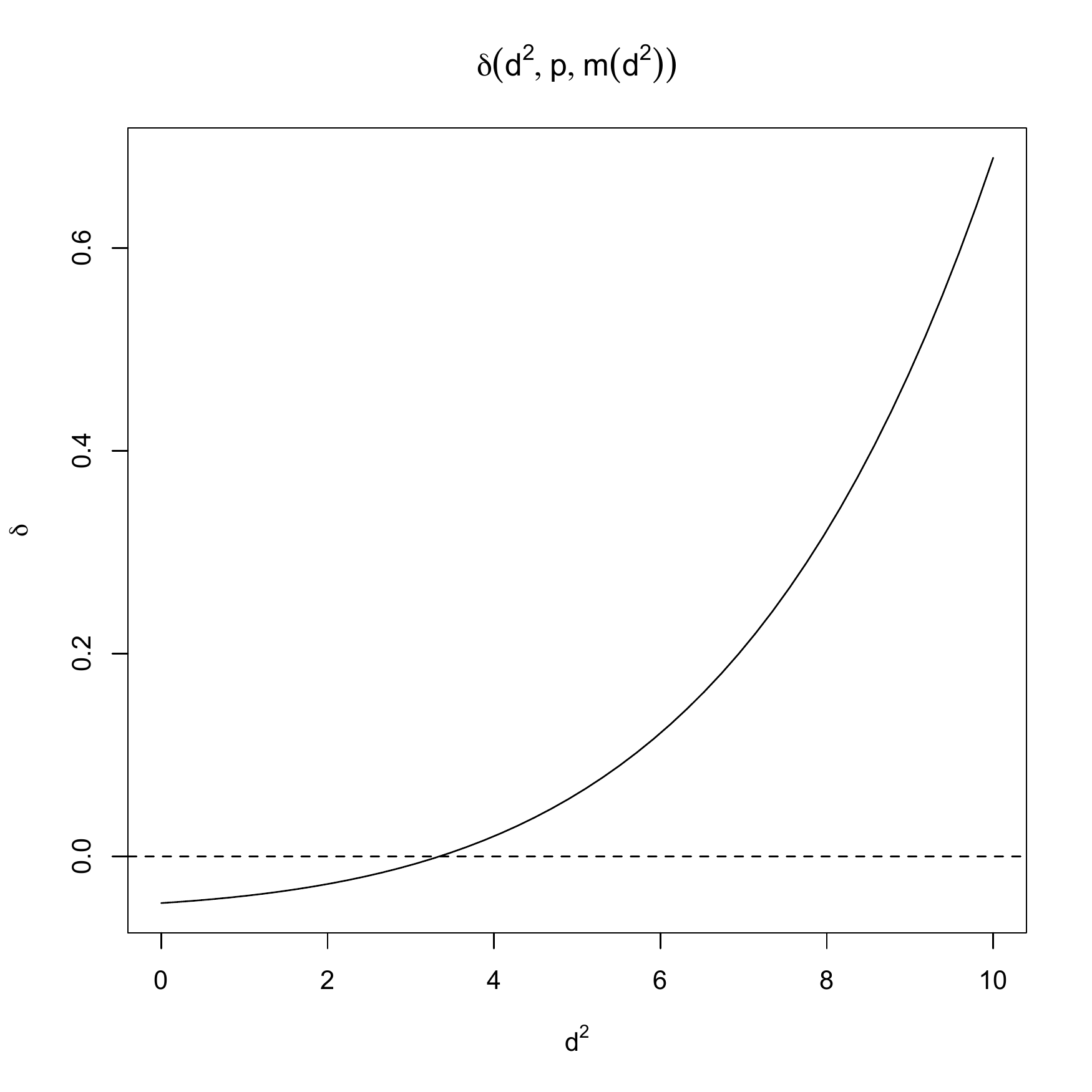}
\includegraphics[width=0.32\textwidth]{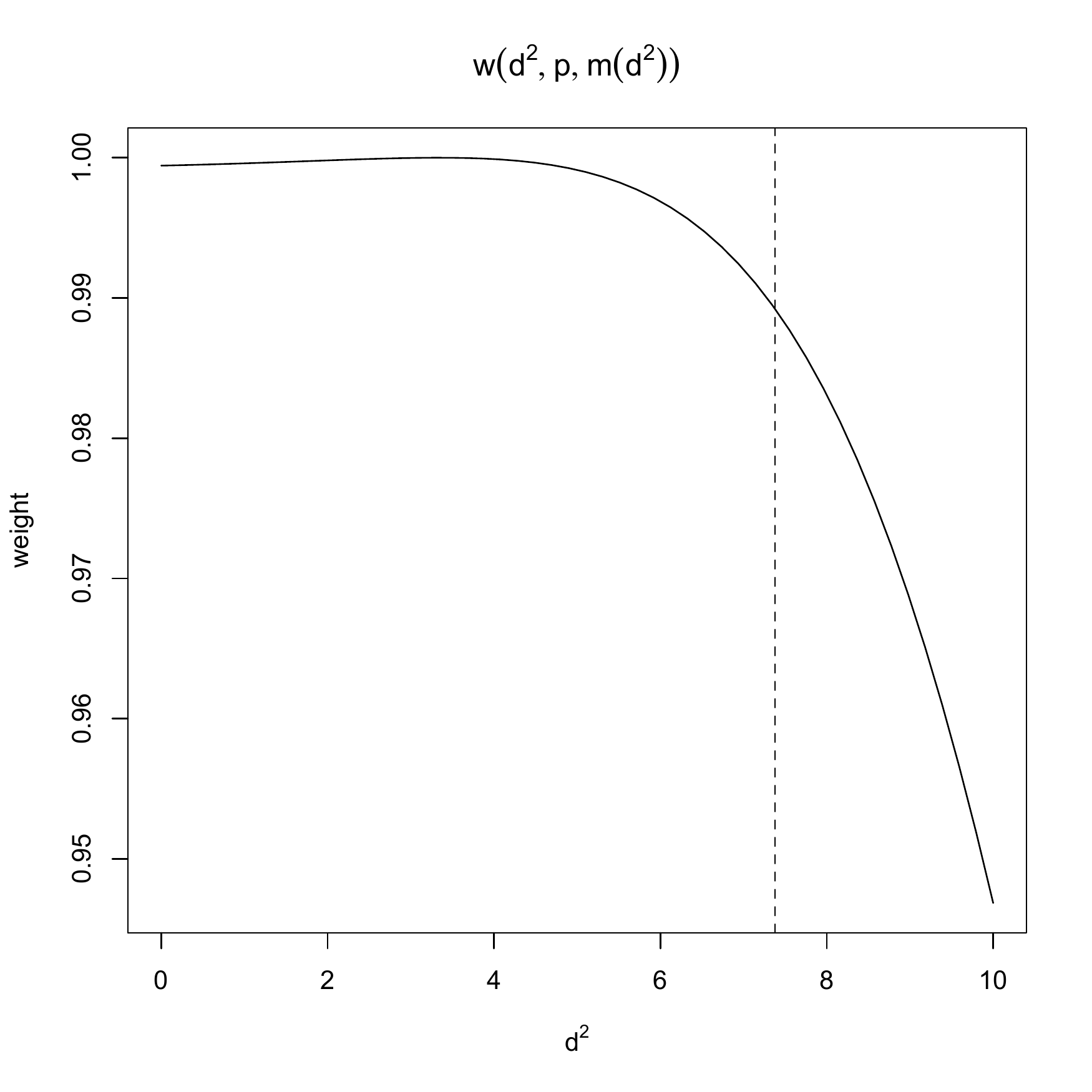}
\caption{Underlying mixture model with its components(left panel). Pearson Residual function (central panel). Weight function (right panel) based on the Hellinger distance RAF.}
\label{deltaw}
\end{figure}

The behavior of the Pearson residual function and the resulting weight function are exemplified in Figure \ref{deltaw}. The true underlying model for the squared distances is assumed to be an $\epsilon$-contaminated model of the form  $m(x)= (1-\epsilon) \chi^2_p(x) + \epsilon \chi^2_p(x,c)$, where the perturbing component is a non-central $\chi^2_p$ distribution with non centrality parameter $c$. 
The mixture model is shown in the left panel, with $p=2, c=5, \epsilon=0.05$. Large squared distances are likely to occur under the contaminating component and are expected to be down-weighted at the $\chi^2_2$ distribution.
The middle panel displays the (asymptotic) Pearson residual function at the $\chi^2_2$ model: actually, 
it takes large values at large distances and, hence, detect a region where outlying distances are likely to occur.
The weight function based on the Hellinger distance RAF is given in the right panel: it clearly decreases 
at large distances. The vertical dashed line in the third panel gives the $0.975$- level quantile of the $\chi^2_2$ distribution: this is the quantile commonly used to declare a large distance and detect outliers in robust multivariate estimation.

The WLE of multivariate location and scatter $(\hat\mu, \hat\Sigma)$ is a weighted mean and weighted covariance matrix with data dependent weights. It is a common practice to consider an unbiased weighted likelihood estimates of the covariance matrix, that can be defined as
$$
\hat\Sigma_u=\frac{\sum_{i=1}^n (y_i-\hat\mu)(y_i-\hat\mu)^{\T}\hat w_i}{\gamma\sum_{i=1}^n \hat w_i}, \quad \gamma=1-\frac{\sum_{i=1}^n \hat w_i^2}{(\sum_{i=1}^n \hat w_i)^2} \ ,
$$
where $\hat w_i=w(\hat d_i^2, \hat\mu, \hat\Sigma_u, \hat M(d_i^2))$, $\hat d_i=d(y_i, \hat\mu,\hat\Sigma_u)$, , $i=1,2,\ldots,n$. Actually, this is the approach currently implemented in the {\tt R} function {\tt cov.wt} to get an unbiased estimate of scatter. When all the weights are equal to one, then $(n-1)$ appears in the denominator.

The computation of $(\hat\mu, \hat\Sigma_u)$ yields an iterative procedure, as illustrated in Algorithm \ref{a1}.
At each iteration, based on the current values $(\hat\mu,\hat\Sigma_u)$ robust distances are obtained. 
Then, their non parametric density estimate is fitted based on the chosen kernel and Pearson residuals and weights are updated.
This algorithm shares the main features of the iterative procedure developed to obtain weighted likelihood estimates in linear regression \citep{agostinelli1998one} and generalized linear models \citep{alqallaf2016robust}.
Actually, in Algorithm \ref{a1}
at each iteration squared distances and their non-parametric density estimate are updated,  whereas the (smoothed) model is held fixed. 

\begin{algorithm}[t]
\caption{WLE based on the Mahalanobis distance} 
\label{a1}
\begin{algorithmic}
\item[] {\bf Initialize $(\hat\mu,\hat\Sigma_u)$}
 \item[] {\bf Calculate squared distances} $$\hat d_i^2=d^2(y_i, \hat\mu,\hat\Sigma_u)$$
\item[] {\bf Evaluate a nonparametric density estimate} $$\hat m_n(d_i^2)$$
\item[] {\bf Compute Pearson residuals} $$\delta(\hat d_i^2; \hat\mu,\hat\Sigma_u, \hat{M}_n) = \frac{\hat m_n(d_i^2) }{m^*(d_i^2)}-1$$
\item[] {\bf Compute weights} $$\hat w_i = \frac{\left[A(\delta(\hat d_i^2, \hat\mu,\hat\Sigma_u, \hat{M}_n)) + 1\right]^+}{\delta(d_i^2,\hat\mu,\hat\Sigma_u, \hat{M}_n) + 1} $$
\item[] {\bf Update $(\hat\mu,\hat\Sigma_u)$}
\end{algorithmic}
\end{algorithm}

Some care is needed in the construction of the kernel density estimate $\hat m_n(d_i^2)$ since it is not expected to allocate any weight before zero otherwise it will be biased at the boundary \citep{karunamuni2005boundary}. 
In the development of Algorithm \ref{a1} four methods are suggested to come through this issue. 
The first three are designed to obtain an unbiased kernel density estimate over $(0,\infty)$, whereas the fourth is based on the distribution of log-transformed squared distances, moving the problem over the whole real line.

\begin{enumerate}
\item The reflection technique \citep{silverman1986density}
is based on data augmentation by adding the reflections of all the points in the boundary. Then, it is possible to implement any method originally designed for the whole real line. A reflection kernel can be defines as follows
$$
k(y; t, h)=\frac{1}{h}k\left( \frac{y-t}{h}\right)+\frac{1}{h}k\left( \frac{y+t}{h}\right)
$$
where $k(\cdot)$ is a symmetric and differentiable probability density: the 
reflection of a normal density leads to a folded normal kernel.
\item A kernel density estimate over $(0,\infty)$ can be also obtained by first log-transforming the squared distances, fitting a non parametric density estimate over the whole real line, i.e. $\hat m_n(\log d^2)$, and then back-transforming the fitted density to $(0,\infty)$, i.e.  $\hat m_n(d^2)=\frac{1}{d^2}\hat m_n(e^{\log d^2})$, \citep{bowman1997applied}. The corresponding smoothed model can be obtained according to the same scheme. In this paper, we make use of the code available from the {\tt R}-package {\tt sm}.
\item The Gamma kernel \citep{chen2000probability, jones2007kernel}
$$
k(y;t,h)=\Gamma(y; t/h+1,h)
$$
where $\Gamma(t; a,b)$ denotes the probability density function of a Gamma variate with shape parameter $a$ and scale parameter $b$. This is an appealing alternative that does not involve any data transformation. The role of $y$ and $t$ may also be switched.

\item Log-transformed squared distances are distributed according to a $\log\chi^2_p$ model
 whose probability density function is
$$
p(x;p)=\frac{1}{2^{p/2}\Gamma(p/2)}\textrm{exp}\left[ \frac{1}{2}(px-\textrm{exp}(x))\right], \  x\in\Real \ .
$$
Then, Pearson residuals and weights can be evaluated on this new scale by comparing the fitted kernel density based on log-transformed squared distances with the $\log\chi^2_p$ distribution.
\end{enumerate}

The smoothing parameter $h$ indexing the kernel function $k(\cdot;\cdot)$ involved in the construction of the Pearson residuals regulates the robustness/efficiency trade-off of the weighted likelihood methodology. 
An appealing feature is that $h$ may be set independently from the scale of the model.  Its value may be determined in order to achieve a fixed expected downweighting level \citep{markatou1998}, 
that is the expected number of observations that will be deleted under the specified model, a fixed asymptotic weight for a given point mass contamination $\epsilon$ at the outlying point $y$ \citep{agostinelli2001test},
but also in an adaptive fashion by monitoring the empirical downweighting level $(1-\bar \omega)$, with $\bar\omega=n^{-1}\sum_{i=1}^n \hat w_i$ \citep{greco2016weighted}.

Algorithm \ref{a1} may be initialized by drawing a large number of random subsets of fixed dimension $(p+1)$.
The sample mean and covariance matrix are evaluated over each subsample and used as starting values \citep{markatou1998}. 
A deterministic solution to set initial values can be also implemented, stemming from that described in \cite{hubert2012deterministic}.
Strategies to select the best root are given in \cite{markatou1998} and \cite{agostinelli2006}.

\section{The distribution of robust distances}
The availability of robust estimates of location and scatter allows to activate some procedures designed to identify multivariate outliers. Actually, the use of robust estimates in place of the sample vector mean and covariance matrix avoids {\it masking} and {\it swamping} effects in outlier detection: there is masking whenever an outlier is not detected because of the presence of similar outliers, swamping when a genuine observation is flagged as an outlier.

The problem of outlier detection consists in
testing the $n$ null hypotheses that each data point is a realization of a multivariate normal distribution $N_p(\mu,\Sigma)$. 
The detection rule will depend on the (asymptotic) distribution of the squared robust distances $d^2(y;\hat\mu,\hat\Sigma_u)$.
A common approach to define cut-off values to flag outliers is based on the $\chi^2_p$ distribution to approximate the distribution of squared robust distances. A more accurate distributional result may be used after the computation of the MCD estimator \citep{cerioli2010multivariate}, but not in the case of M-type estimation. A rule of thumb is based on the $0.975$-level quantile of the reference distribution.
The outliers detection process could also be designed to take into account multiplicity arguments in the simultaneous testing of all the $n$ data points.  For instance, 
cut-off values can be based on a $(1-\gamma)$-level quantile such that
the simultaneous testing of all the data points corresponds to a global nominal level $\alpha$, that is
$\gamma=1-(1-\alpha)^{1/n}$  \citep{cerioli2010multivariate} or by controlling the overall level of the simultaneous testing procedure by the False Discovery Rate \citep{cerioli2011error}.

Here, we state a result concerning the distribution of robust distances at the postulated multivariate normal model based on the WLE, that resembles, asymptotically, the classical one on the Mahalanobis distance evaluated at the unbiased MLE  \citep{gnanadesikan1972robust}.
\\

{\bf Proposition 1} Let $\hat\theta=(\hat\mu,\hat\Sigma_u)$ be the unbiased WLE of multivariate location and scatter. Assume that: (i) the model is correctly specified, that is $m(y)=m(y;\theta_0)$ for some $\theta_0=(\mu_0,\Sigma_0) \in \Theta$; (ii) $\hat{\theta}$ is a consistent estimator of $\theta_0$; (iii) $\sup_y \left| w (y;\hat{\theta}, \hat{M}_n) - 1 \right| \stackrel{p}{\longrightarrow} 0$. Then
\begin{equation}
\label{prop1}
d^2(Y_i, \hat\mu,\hat\Sigma_u) \stackrel{d}{\rightarrow} \frac{(n-1)^2}{n} \textrm{Beta}\left( \frac{p}{2}, \frac{n-p}{2}\right) \ .
\end{equation}
\\

The proof is given in the Appendix.  It follows the guidelines of the classical proof based on the unbiased MLE that has been revised in \cite{ververidis}.
It is worth to stress that the same result does not hold for M-type estimation since Huber and Tukey's bisquare weights do not share the asymptotic behavior of the weights in (\ref{weight}) at the postulated model. A close result has been established in the case of the MCD \citep{cerioli2010multivariate}.

\section{Numerical studies}
In this section we investigate the finite sample behavior of the newly proposed WLE of multivariate location and scatter through some numerical studies. The strategies outlined in Section 3 to compute Pearson residuals and the corresponding weights are all considered: folded normal kernel (WLEa), log and back transform (WLEb), log transform with $\log\chi^2_p$ (WLEc), gamma kernel (WLEd). The weights are based on the Hellinger distance RAF.
The multivariate WLE has been also compared with the deterministic MCD and the S-estimator (with Rocke type weights, that has been designed to work properly for large dimensions), evaluated by using the functions from the {\tt R} package {\tt rrcov}.
The WLE runs on a deterministic algorithm \citep{hubert2012deterministic}, that is based on six initial solutions. For each starting point, Pearson residuals are evaluated and the solution with the lowest fitted probability $$\textrm{Pr}_{\hat\theta}\left[\delta(Y; \hat\theta, \hat M_n)<-0.95\right]$$ is selected \citep{agostinelli2006}. Then, this initial proposal is iteratively updated according to Algorithm 1.

Several combinations of $(n,p)$  have been taken into account. 
Data have been generated according to a multivariate normal $N_p(0, I)$ with uncorrelated components and unit variance with point mass contamination, that is a percentage $\epsilon$ of outliers is driven by a multivariate normal model $N_p(ka, \delta I)$ with $k=1,2, 3, \ldots, K$ and $\delta=0.01$ ($k=0$ gives the uncontaminated scenario). When $p\leq 10$, contamination is designed to affect all dimensions, $a=(1,1,\ldots,1)^{\T}$, wheres when $p>10$, outliers only contaminate the first five dimensions $a=(1,1,1,1,1,0,\ldots,0)^{\T}$.
We show results corresponding to a contamination level $\epsilon=20\%$ and two data configurations: in the first $n=100$ and $p=10$, in the second $n=500$ and $p=50$. The most distant point mass contamination has been located at $K=5$ in the first case and at $K=10$ in the second one.
All numerical studies are based on 1000 Monte Carlo trials.

The following performance measures were considered:
\begin{enumerate}
\item $||\hat\mu||$
\item $\log\frac{\textrm{trace}(\hat\Sigma_u)}{p}$
\item $\log_{10}\textrm{cond}(\hat\Sigma_u)$
\item $||\hat\Sigma_{u_{jj}}-1||, \ j=1,2,\ldots,p$
\item $||\hat\Sigma_{u_{jh}}||, \ j,h=1,2,\ldots,p, j\neq h$
\item computational time (in seconds on a 3,4 GHz Intel Core i5).
\end{enumerate}
All of them are expected to be as close as possible to zero.
  
\begin{figure}[th]
\centering
\includegraphics[width=0.49\textwidth]{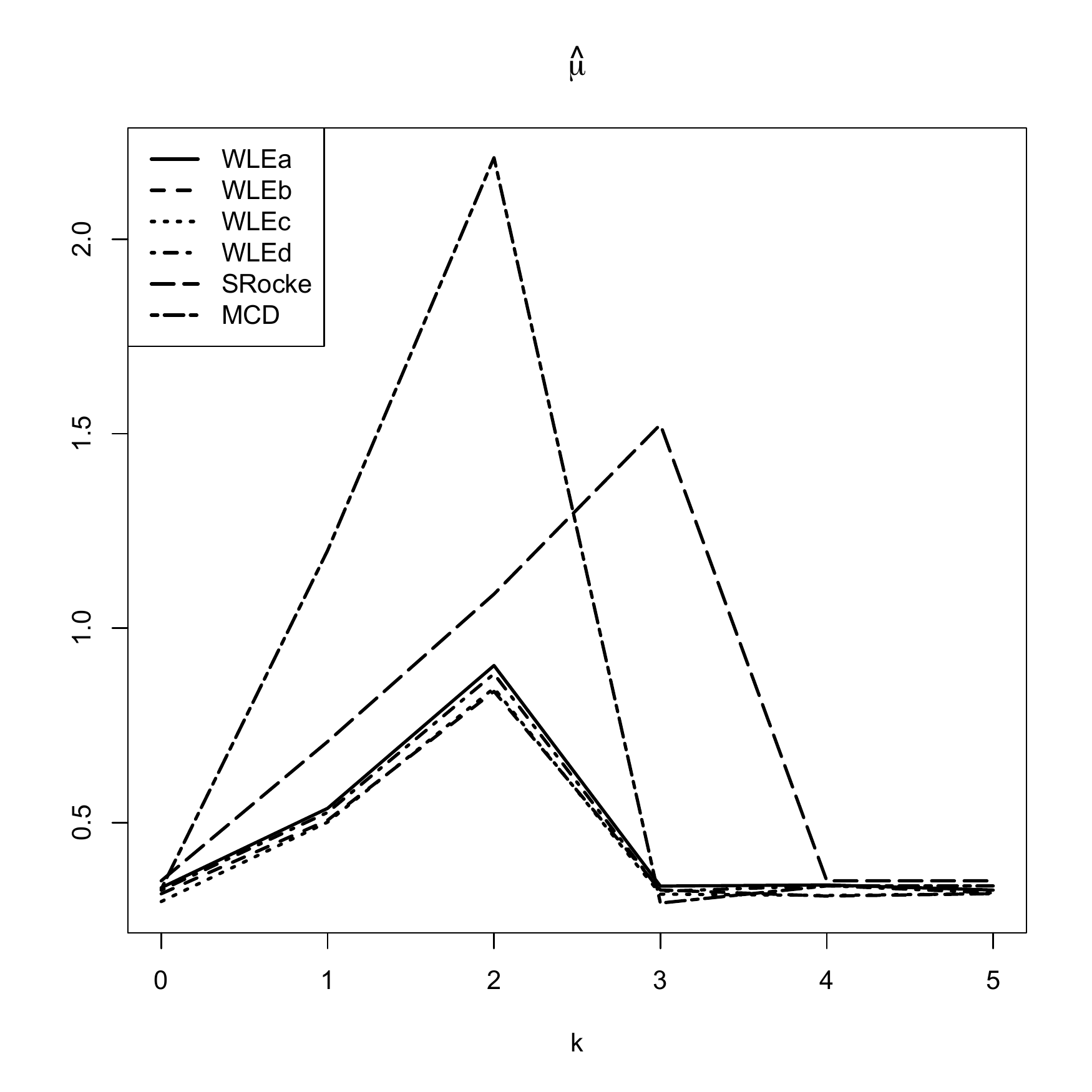}
\includegraphics[width=0.49\textwidth]{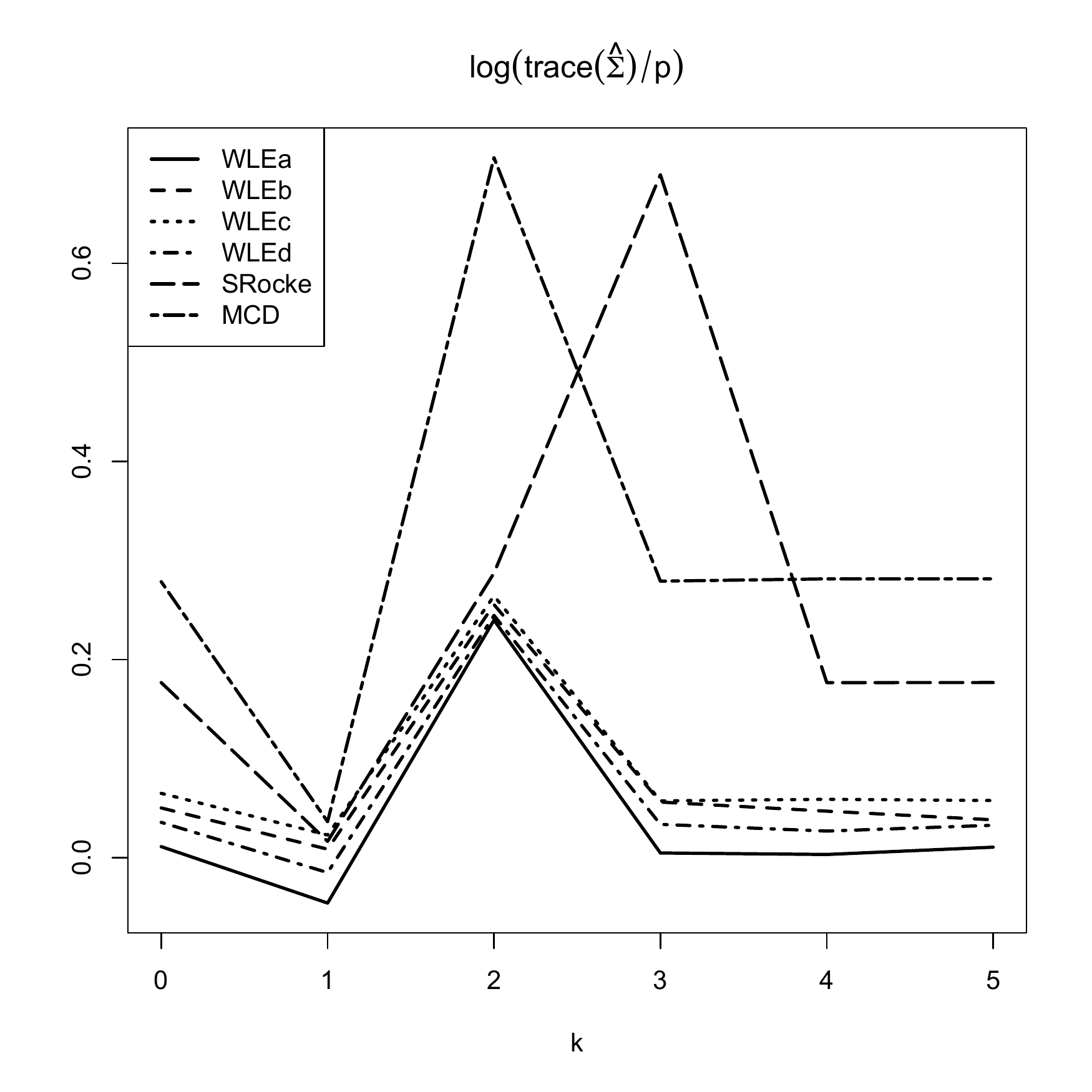}
\includegraphics[width=0.49\textwidth]{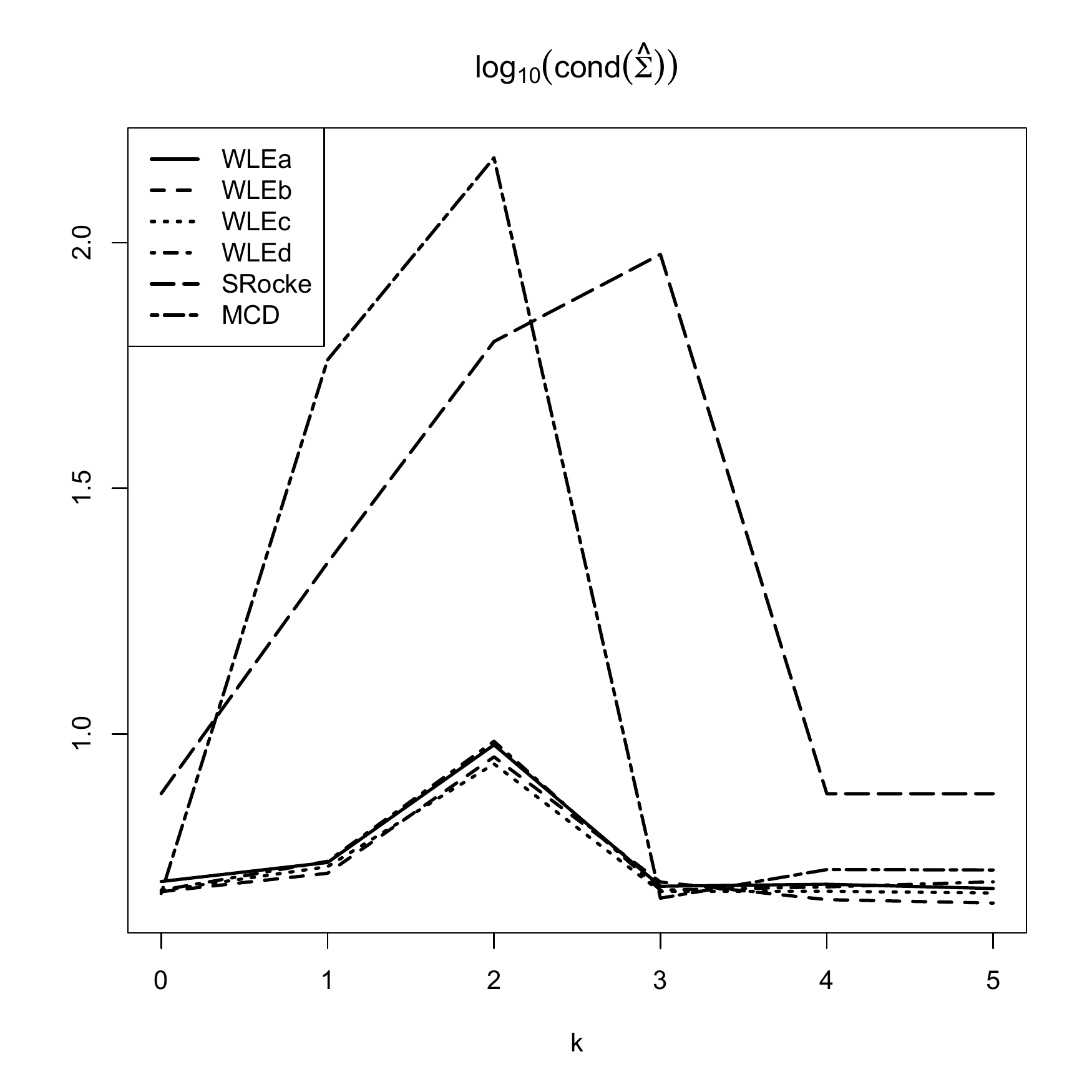}
\includegraphics[width=0.49\textwidth]{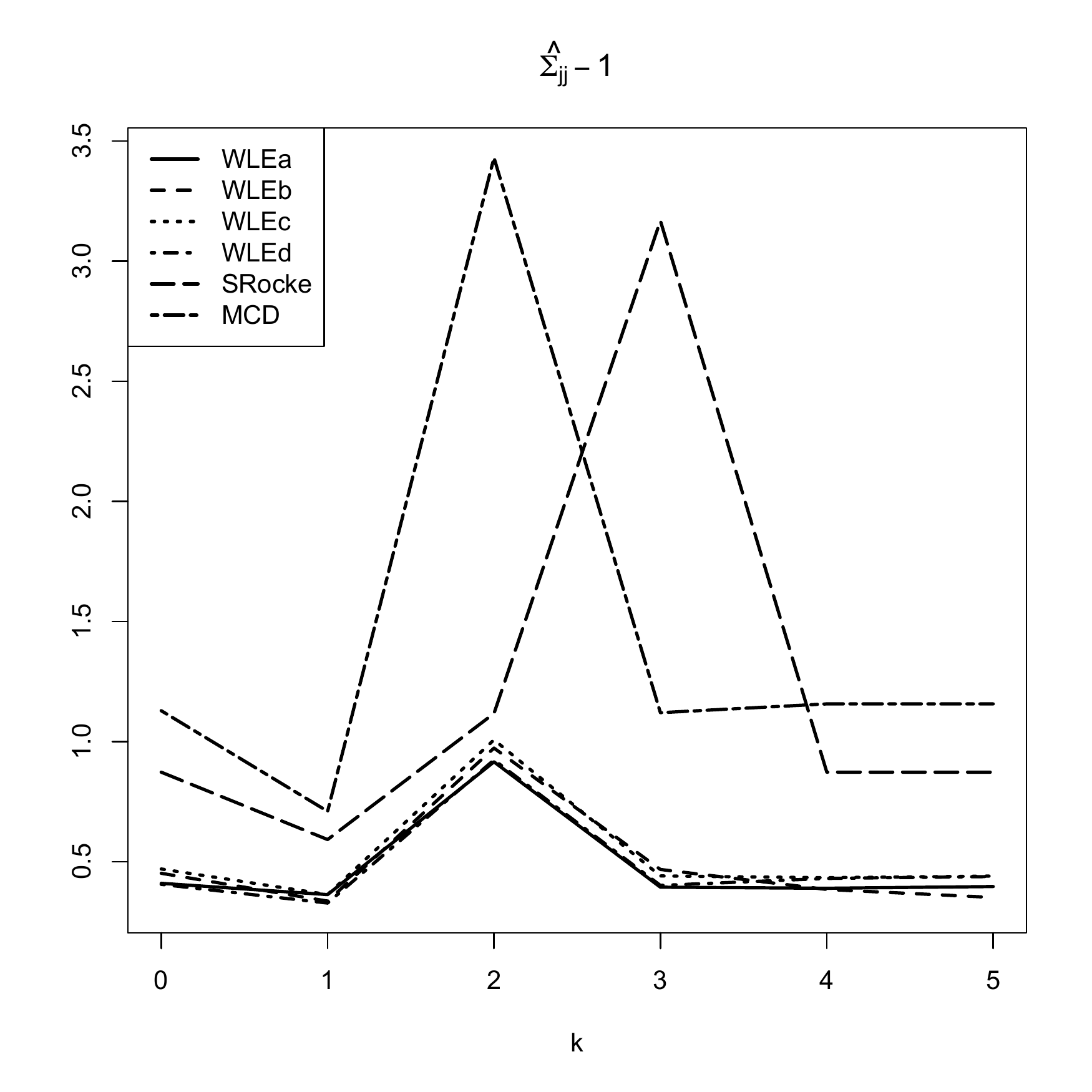}
\includegraphics[width=0.49\textwidth]{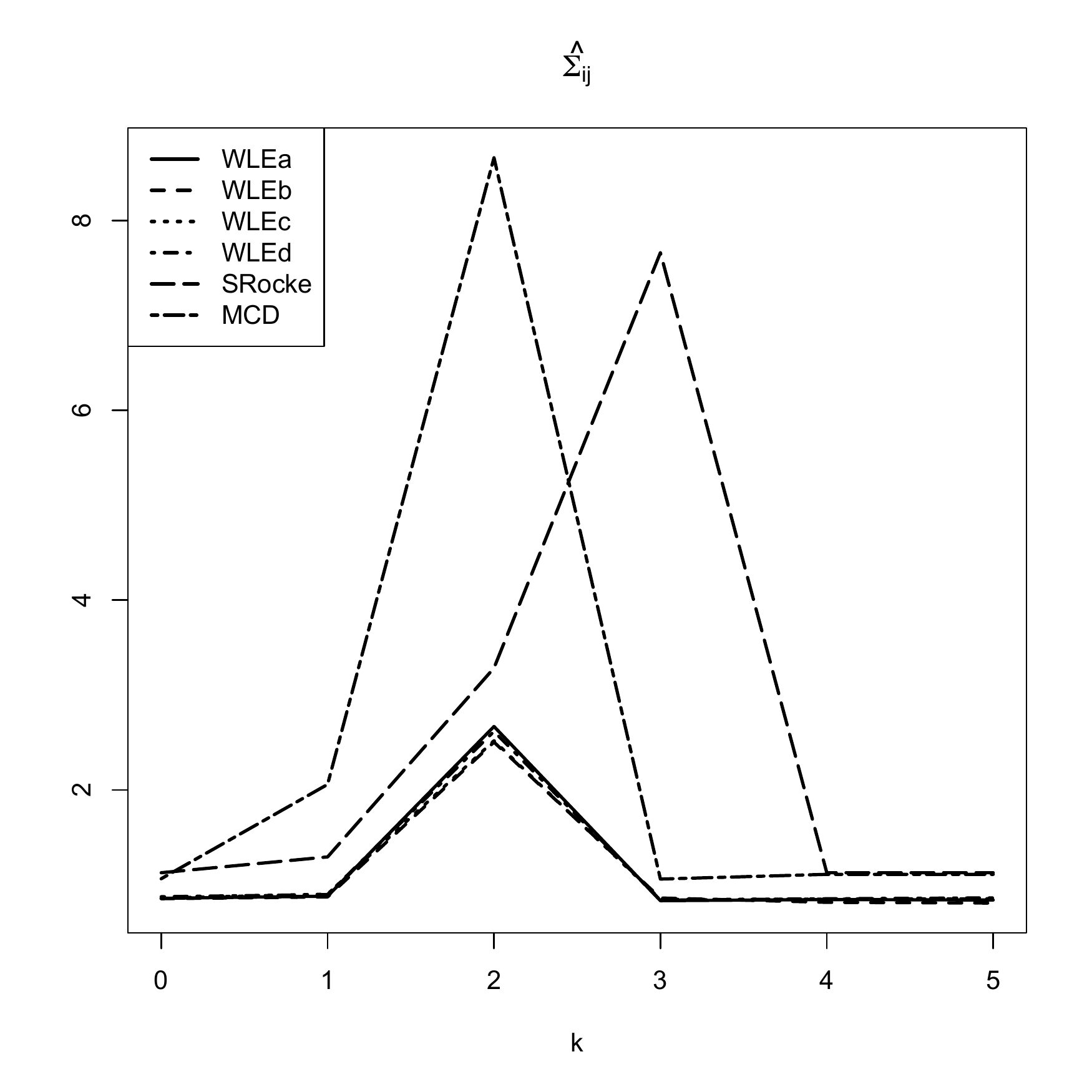}
\includegraphics[width=0.49\textwidth]{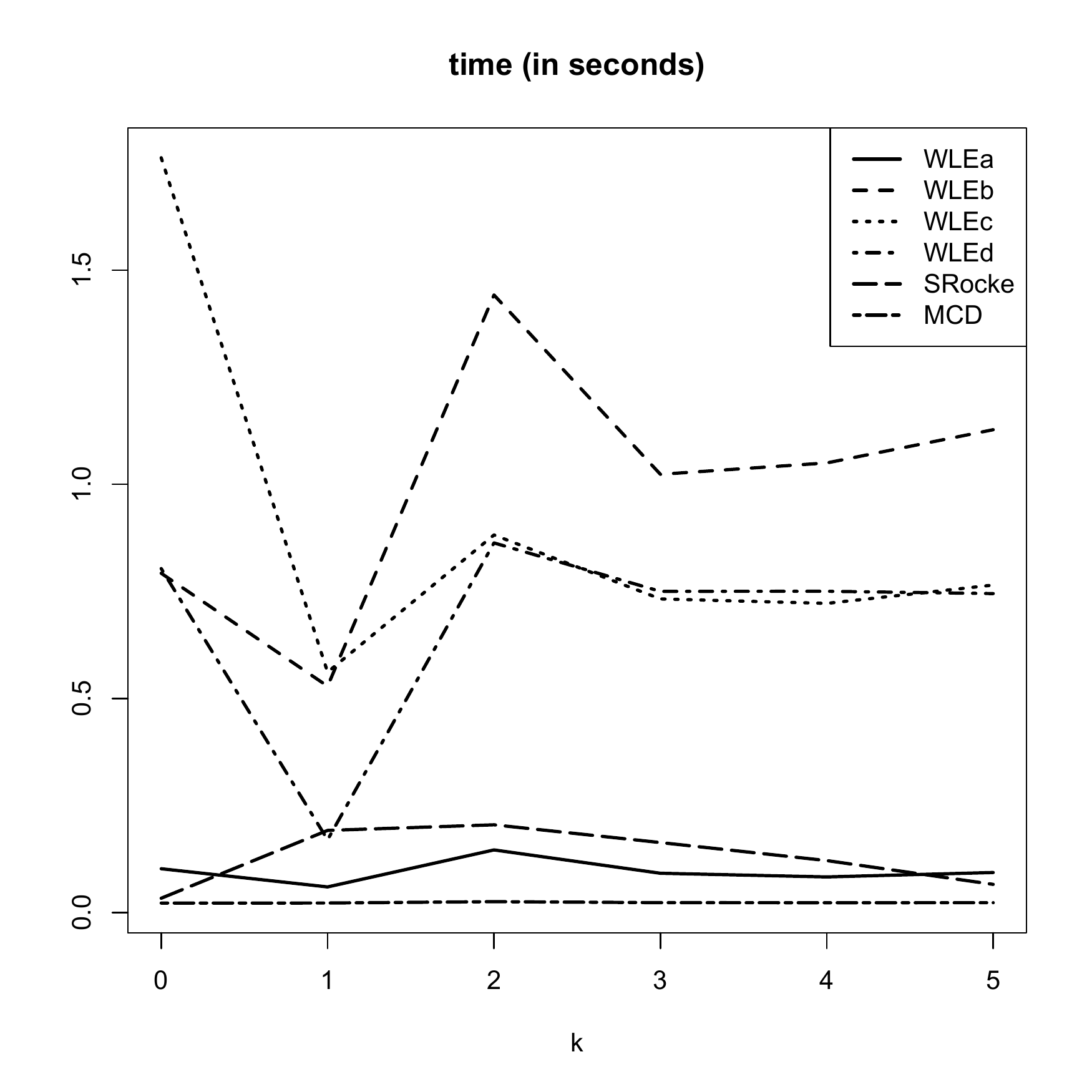}
\caption{Numerical study: average performance measures when $n=100$, $p=10$, $k=0,1,2,3,4,5$, $\epsilon=20\%$, for WLEa, WLEb, WLEc, WLEd, SRocke and MCD}
\label{sim1}
\end{figure}

\begin{figure}[th]
\centering
\includegraphics[width=0.49\textwidth]{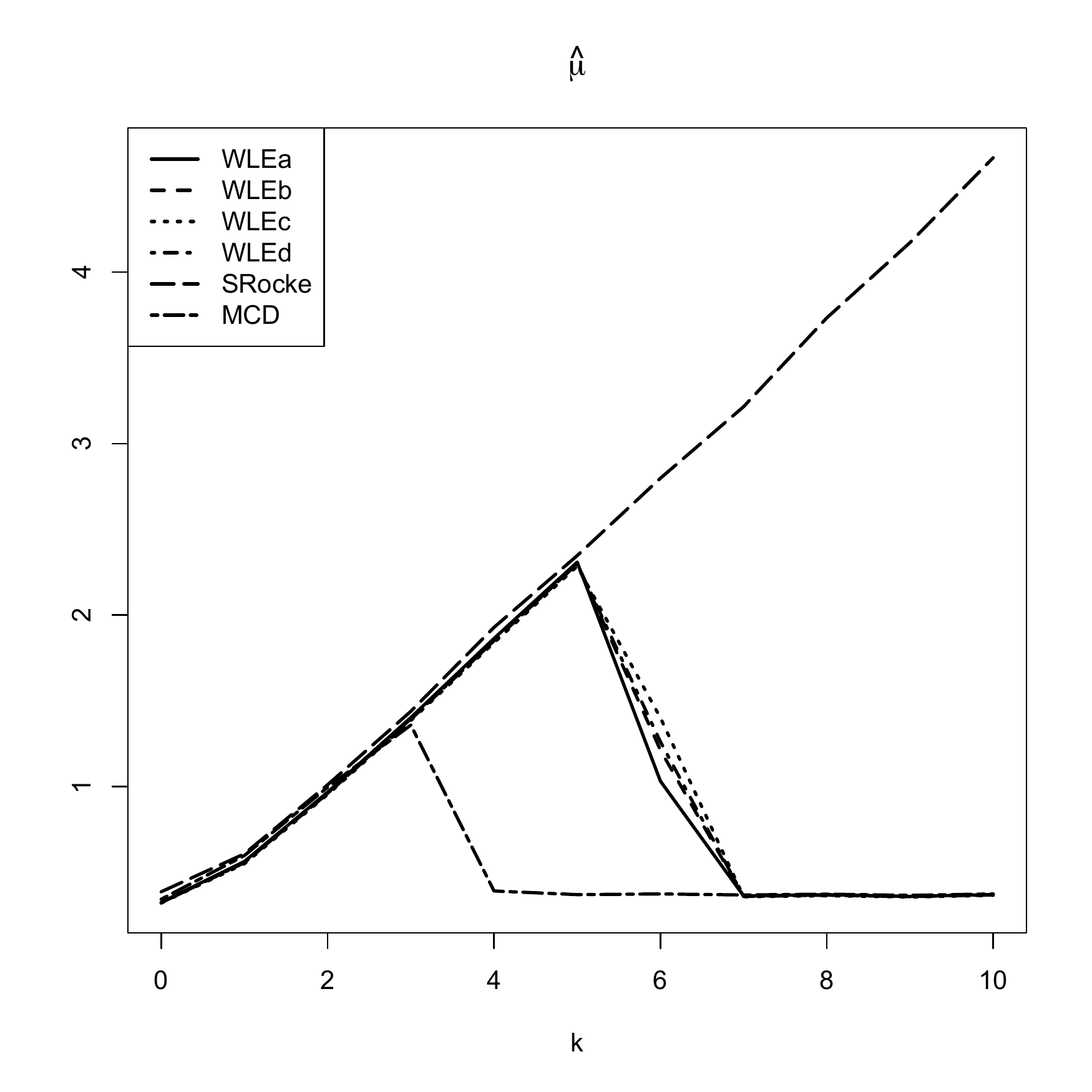}
\includegraphics[width=0.49\textwidth]{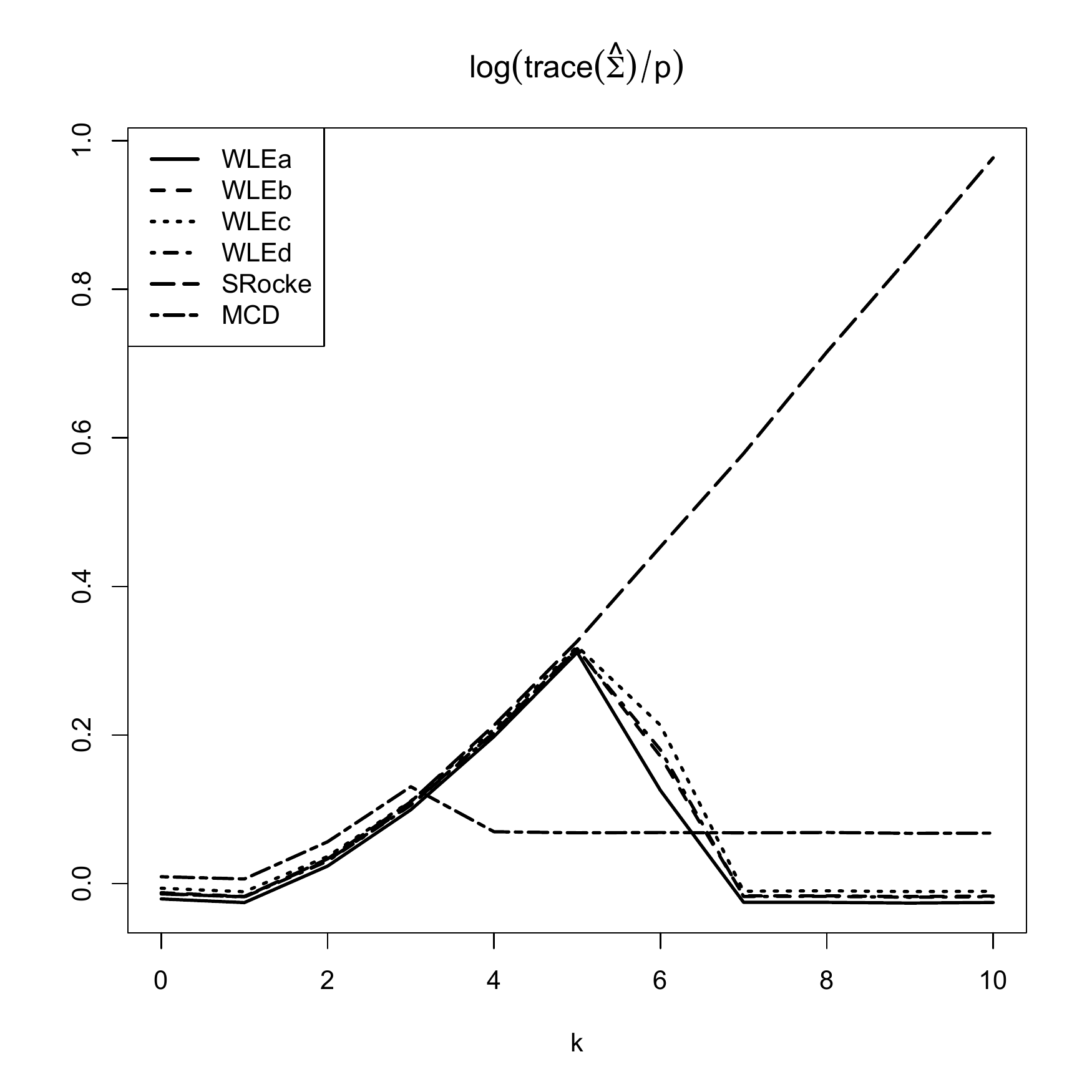}
\includegraphics[width=0.49\textwidth]{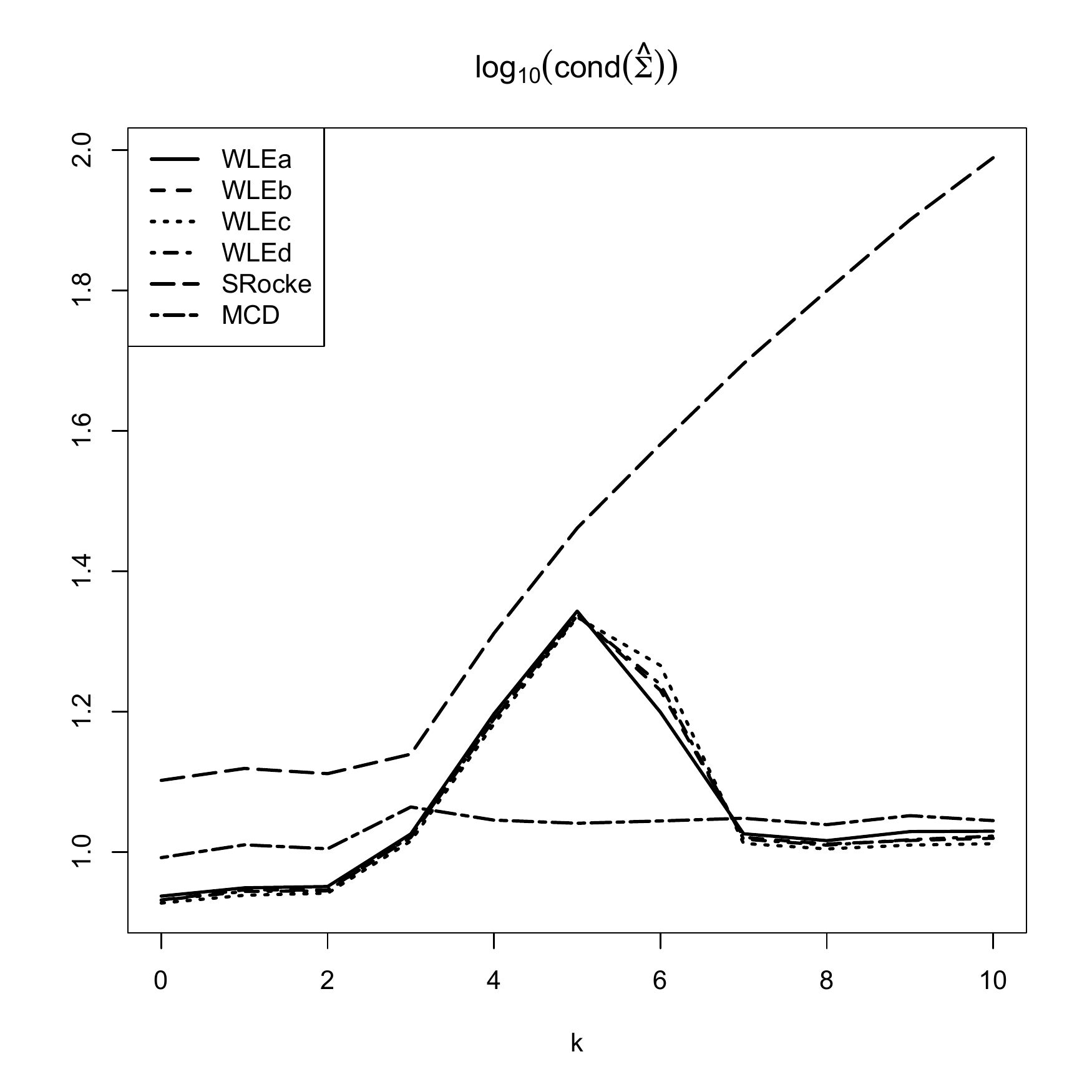}
\includegraphics[width=0.49\textwidth]{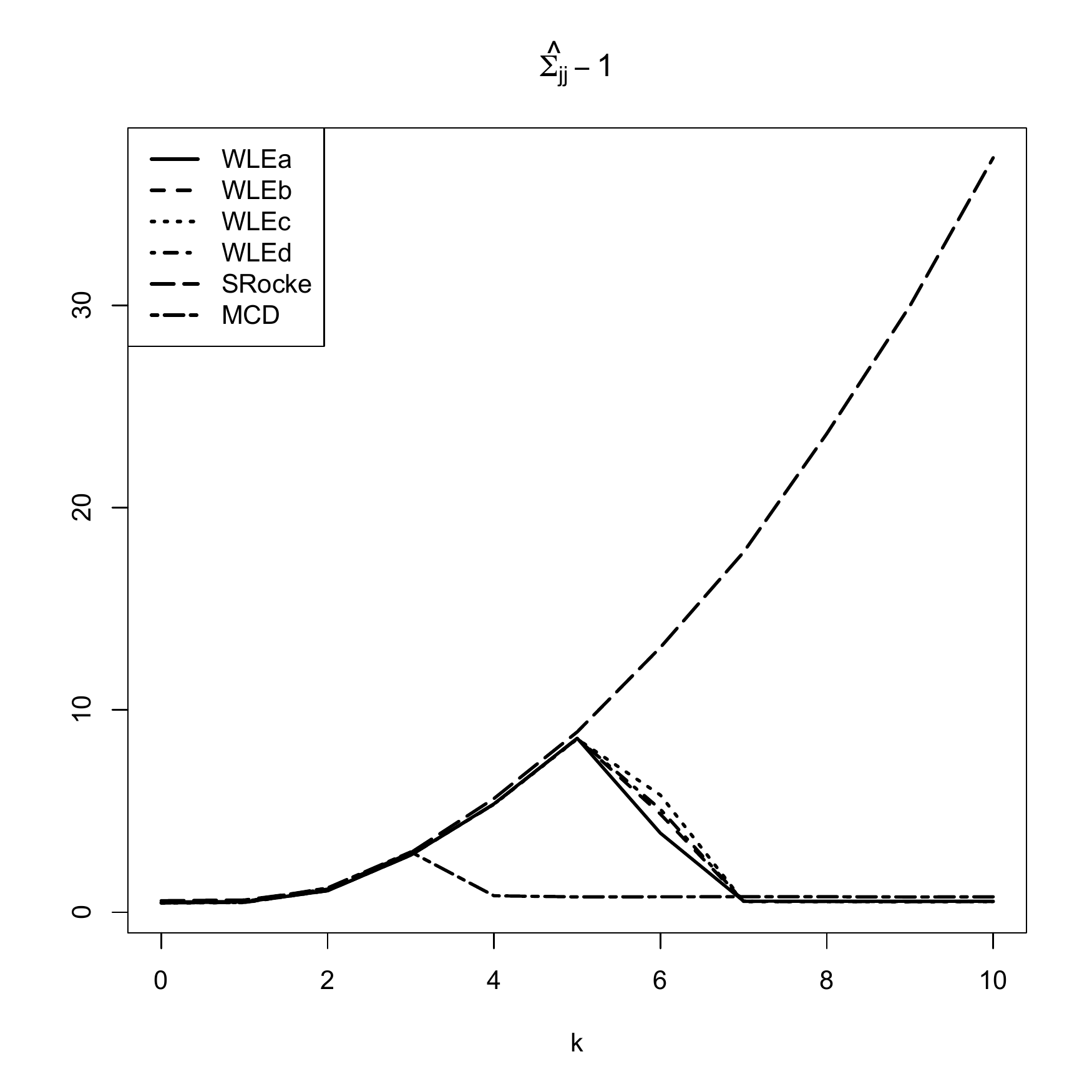}
\includegraphics[width=0.49\textwidth]{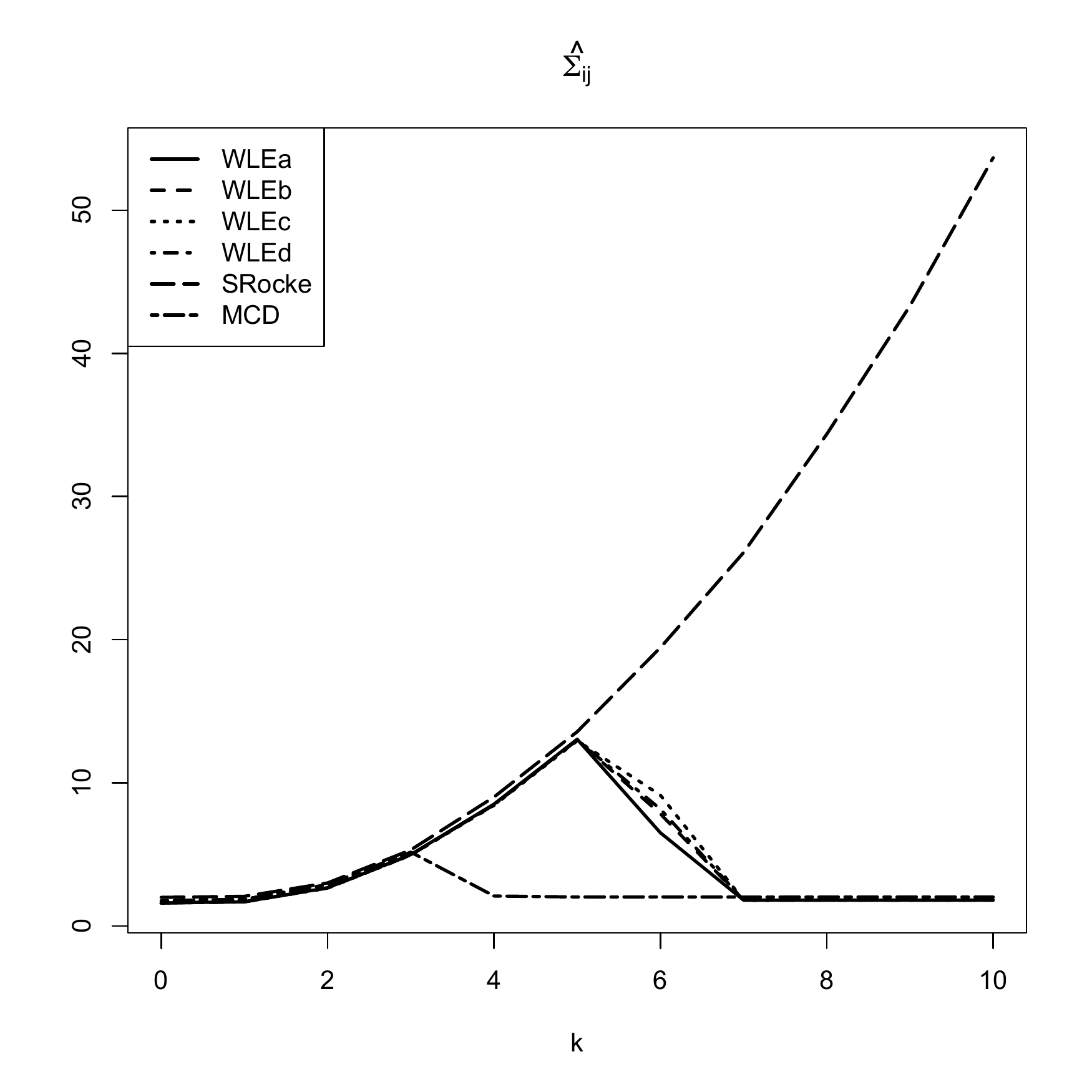}
\includegraphics[width=0.49\textwidth]{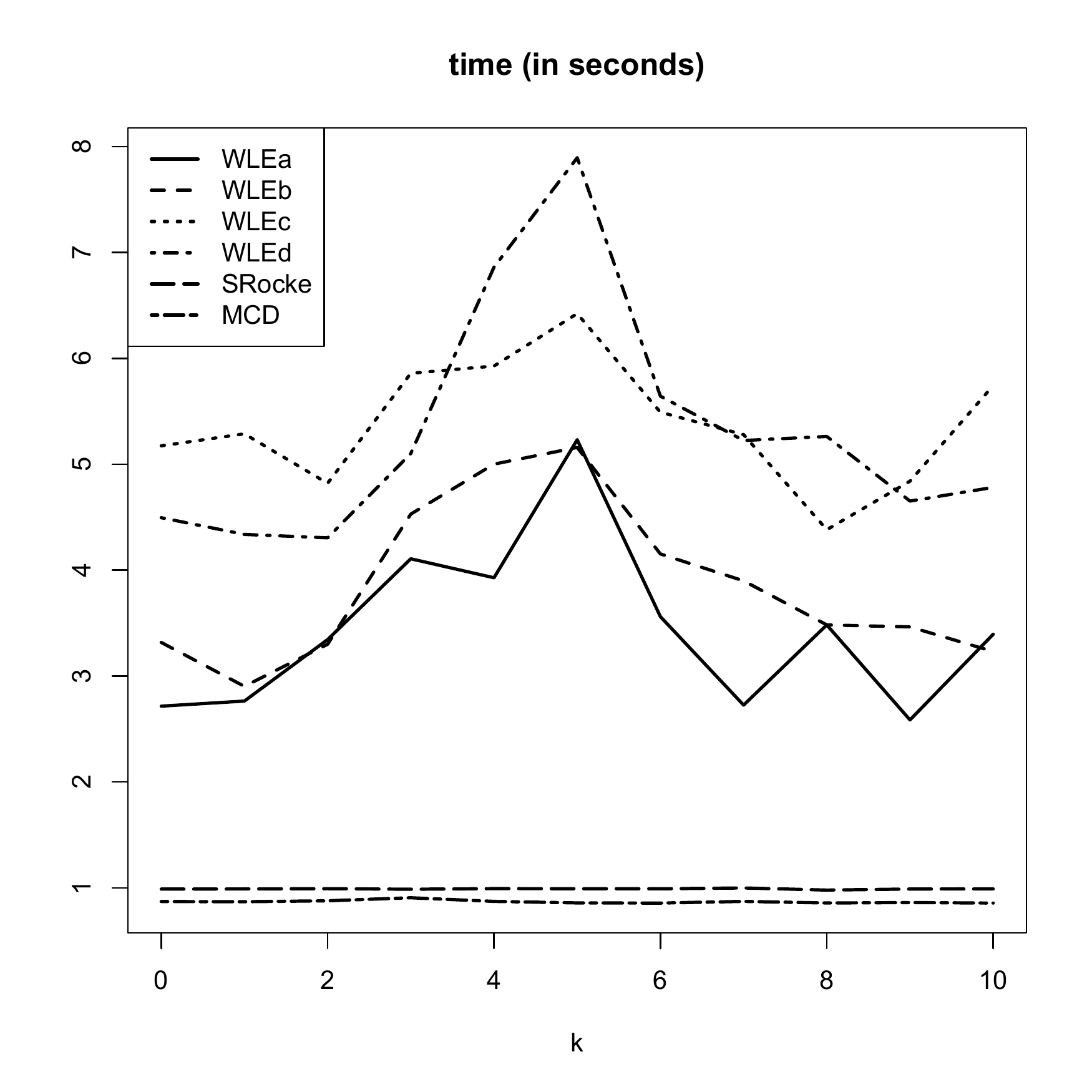}
\caption{Numerical study: average performance measures when $n=500$, $p=50$, $k=0,1,2,3,\ldots,10$, $\epsilon=20\%$, for WLEa, WLEb, WLEc, WLEd, SRocke and MCD.}
\label{sim2}
\end{figure}

Figure \ref{sim1} displays the average performance measures for  $n=100$, $p=10$.
The WLE provides very accurate results and an appealing behavior compared to the S-estimator and the MCD, whatever the chosen weighting scheme, both for location and covariance estimation. 
It is worth noting that all the estimators provide less accurate results when outliers are not located at large distances.
The WLE becomes less accurate at $k=2$ only and outperforms both the S-estimator and the MCD. The former suffers from contamination still at $k=3$, whereas the latter exhibits the desired performance at $k=2$, as well, but the WLE is to be preferred.

The results corresponding to the second data configuration, with $n=500$ and $p=50$ are showed in Figure \ref{sim2}.
In the same fashion as above,  
contamination is such that all estimators may  suffer from lack of robustness when outliers are not located at large distances.  
The WLE still exhibits a satisfactory behavior. All the estimators share the same features until $k=3$. After that, the MCD shows the desired {\it redescending} behavior, whereas the WLE
is no more affected by contamination at $k=5$ and the S-estimator does not protect inference from contamination for all considered values of $k$. The fact that the WLE performs remarkably better than the S-estimator is a noticeable result, in that both of them fall in the general category of soft-trimming estimators, as stated in the Introduction. 

The sixth panels in both Figures show the computational time. One needs to keep in mind that the comparison
with the  MCD and S- estimators is unfair, since the WLE is still based on an unoptimized {\tt R} code, that will be soon available from package {\tt wle}.
Actually, computational time for the WLE remains in a feasible range, even when $p=50$.
In the first scenario, the use of folded normal leads to save computational time w.r.t. the other kernels, especially for larger $k$. With growing dimensionality, the reflection kernel is still to be preferred but the procedure based on log and back transform is also competitive.

\section{Real data examples}
In this section we provide some real data examples concerning multivariate estimation of location and scatter, outlier detection, principal component analysis and discriminant analysis. The proposed weighted likelihood methodology is also compared with other popular robust multivariate tools.

\subsection{Multivariate estimation}
\begin{figure}[t]
\centering
\includegraphics[height=0.45\textheight]{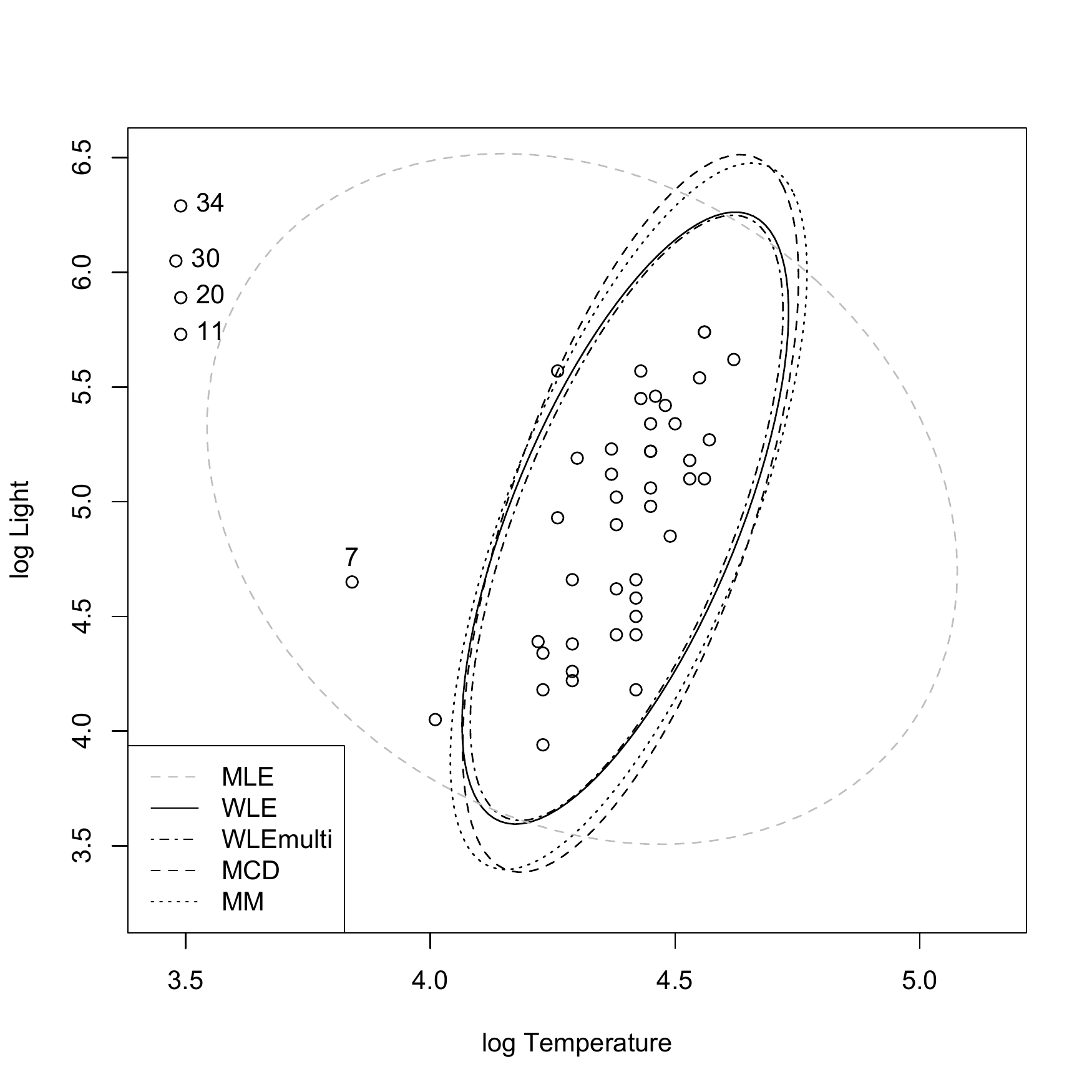}
\caption{StarsCYG data: fitted $97.5\%$ tolerance ellipse based on the WLE, WLEmulti, MCD and MM-estimator.}
\label{stars1}
\end{figure}

The StarsCYG data give the effective temperature at the surface and the light intensity, both on a log scale, of 47 stars in the star cluster CYG OB1. Five stars are clear outliers (the points 11, 20, 30, 34 correspond to giant stars that do not lie on the main sequence) and the point 7 also does not share the correlation structure of the remaining 43 stars.
Figure \ref{stars1} displays 0.975-level tolerance ellipses stemming from the proposed WLE, the WLE based on a bivariate kernel density estimate (WLEmulti),
the MCD (with $50\%$ breakdown point), the MM-estimator (with $50\%$ breakdown point and $95\%$ shape efficiency) and the MLE. The weighted likelihood contours are based on the asymptotic result given in (\ref{prop1}), whereas that stemming from the MCD is based on the distributional result given in \cite{cerioli2010multivariate} and that derived from the MM-estimator is based on the $\chi^2_2$ distribution. The classical tolerance ellipse is also based on the scaled Beta distribution. All the methods we outlined to compute the weights for the WLE gave very similar results and hence, only the one based on log and back transform of distances is shown.
The fitted robust ellipses do not exhibit any significant difference and are all able to catch the correlation structure in the main sequence of stars: the WLE gives a correlation of 0.680, the MCD gives 0.655 and the MM gives 0.691. Moreover, the newly proposed WLE behaves not dissimilarly from the WLE based on the bivariate kernel. On the contrary, the MLE leads to inflated variability and negative correlation.

\begin{figure}[t]
\centering
\includegraphics[height=0.45\textheight]{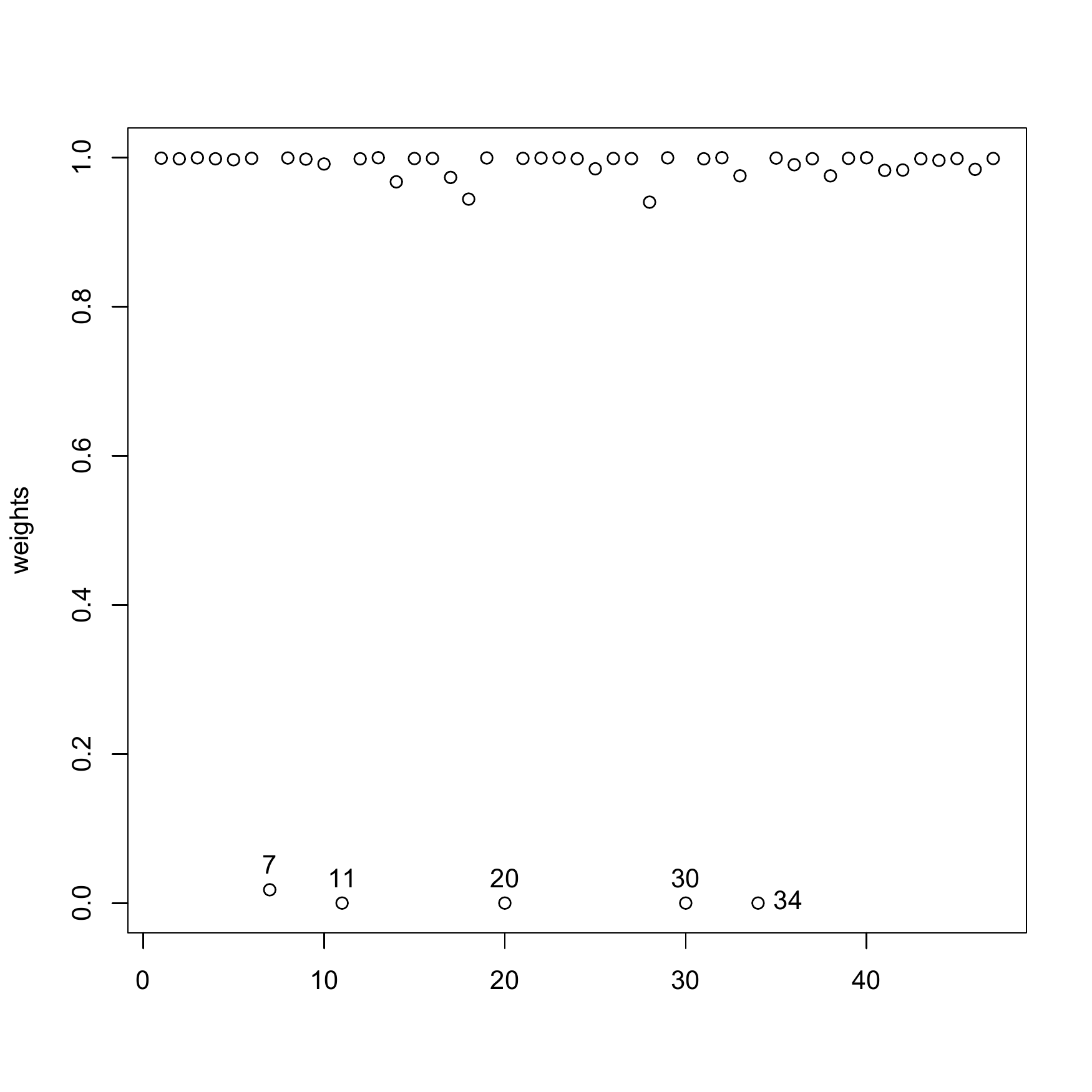}
\caption{StarssCYG data: final robustness weights from WLE.}
\label{stars2}
\end{figure}

Figure \ref{stars2} gives the final robustness weights corresponding to the WLE. All the outliers are given a weight that is very close to zero. The selected smoothing parameter leads to an empirical downweighting level that is about $11\%$ (that is 5 observations over 47).

\subsection{Outliers detection}
\begin{figure}[t]
\centering
\includegraphics[height=0.44\textheight]{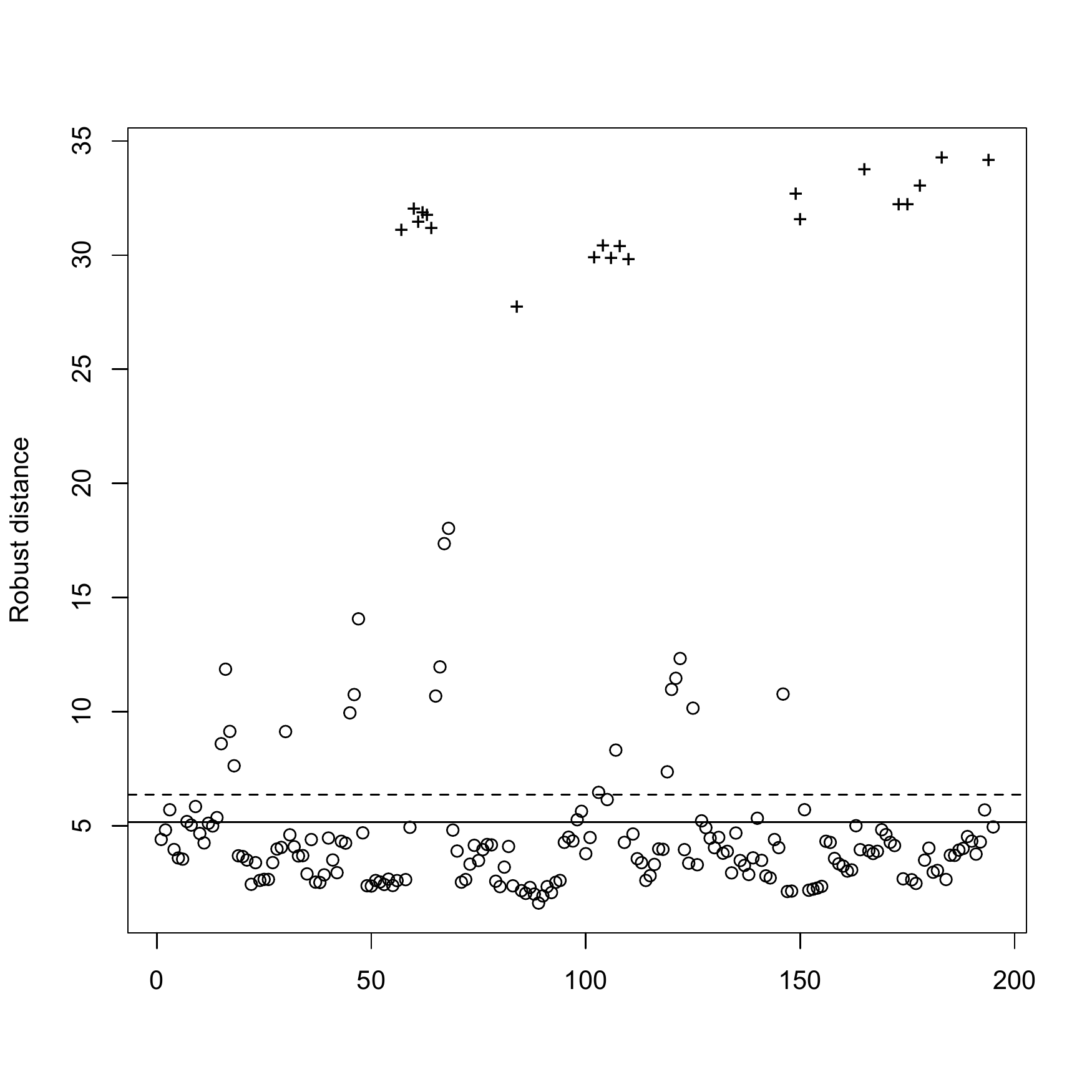}
\caption{Auto data: robust distances based on WLE. Cars running on a diesel engine are denoted by a $+$. The solid line gives the 0.975-level cut-off value, the dashed line gives $(1-\gamma)$-level cut-off value to take into account multiplicity.}
\label{auto1}
\end{figure}

The Auto data give information on technical and insurance characteristics of $n=195$ cars collected in 1985 by the Insurance Institute for Highway Safety, for a total of $p=15$ variables. The car are of two types: running on a gasoline or diesel engine. The are only 20 cars running on diesel that may be identified as outliers w.r.t. the others but several outliers of different nature may arise corresponding, for instance, to cars with peculiar technical features or deserving specific insurance conditions. Figure \ref{auto1} gives the robust distances corresponding to each car stemming from the WLE. 
The solid line gives the cut-off value at the 0.975 level, whereas the dashed line gives the cut-off value at the $(1-\gamma)\approx 0.9998702$  level quantile of the scaled Beta distribution in (\ref{prop1}). The latter threshold
has been computed to take into account multiplicity, in such a way that the simultaneous testing of all the data points correspond to a nominal level 0.025.

\begin{figure}[t]
\centering
\includegraphics[height=0.44\textheight]{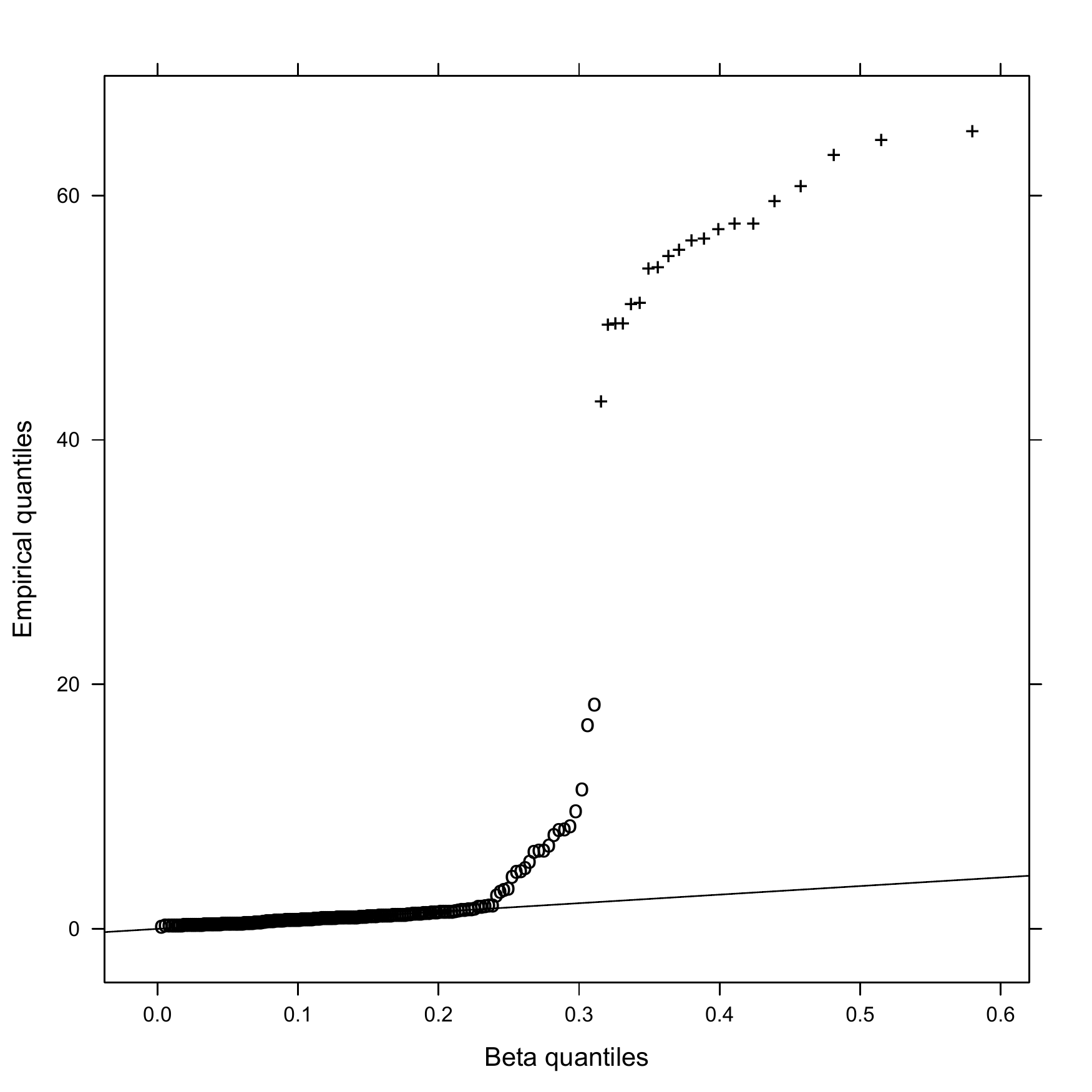}
\caption{Auto data: QQ-plot for robust distances based on WLE. Cars running on a diesel engine are denoted by a $+$.}
\label{auto2}
\end{figure}

\begin{figure}[t]
\centering
\includegraphics[height=0.45\textheight]{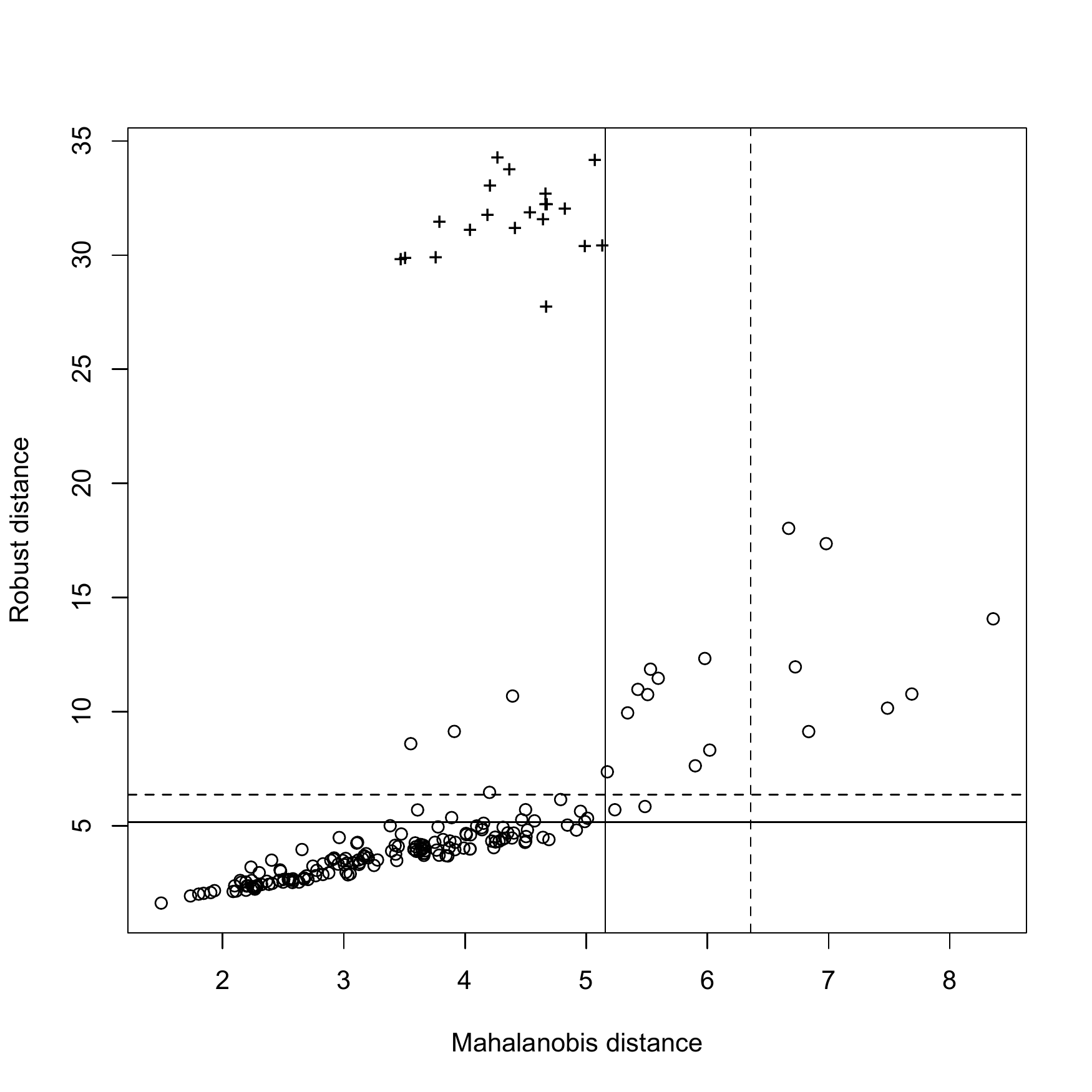}
\caption{Auto data: distance-distance plot based on the WLE. Cars running on a diesel engine are denoted by a $+$.
The solid lines give the 0.975-level cut-off values, the dashed lines give $(1-\gamma)$-level cut-off value to take into account multiplicity.}
\label{auto3}
\end{figure}

The group of cars running on diesel is clearly characterized by the largest distances and is well separated from the remaining cars. The inspection of  Figure \ref{auto1} also unveils some other outlying cars that may exhibit peculiar characteristics. The group of cars running on diesel and the other outliers are clearly spotted by the QQ-plot in Figure \ref{auto2}.
The different nature of the several outliers that we have identified can be investigated further by exploring the distance-distance plot in Figure \ref{auto3}. The robust distances based on the WLE are compared with the classical distances based on the MLE.
An important feature of such plot is that the cut-off values are determined according to the same scaled Beta distribution, hence being the same on both axes. It is worth noting that the group of car running on a diesel engine would not have been detected by looking at the classical distances based on the MLE. The results driven by the use of the MCD, S and MM estimators are very similar. 

\subsection{Principal Component Analysis}
\begin{figure}[t]
\centering
\includegraphics[width=0.49\textwidth]{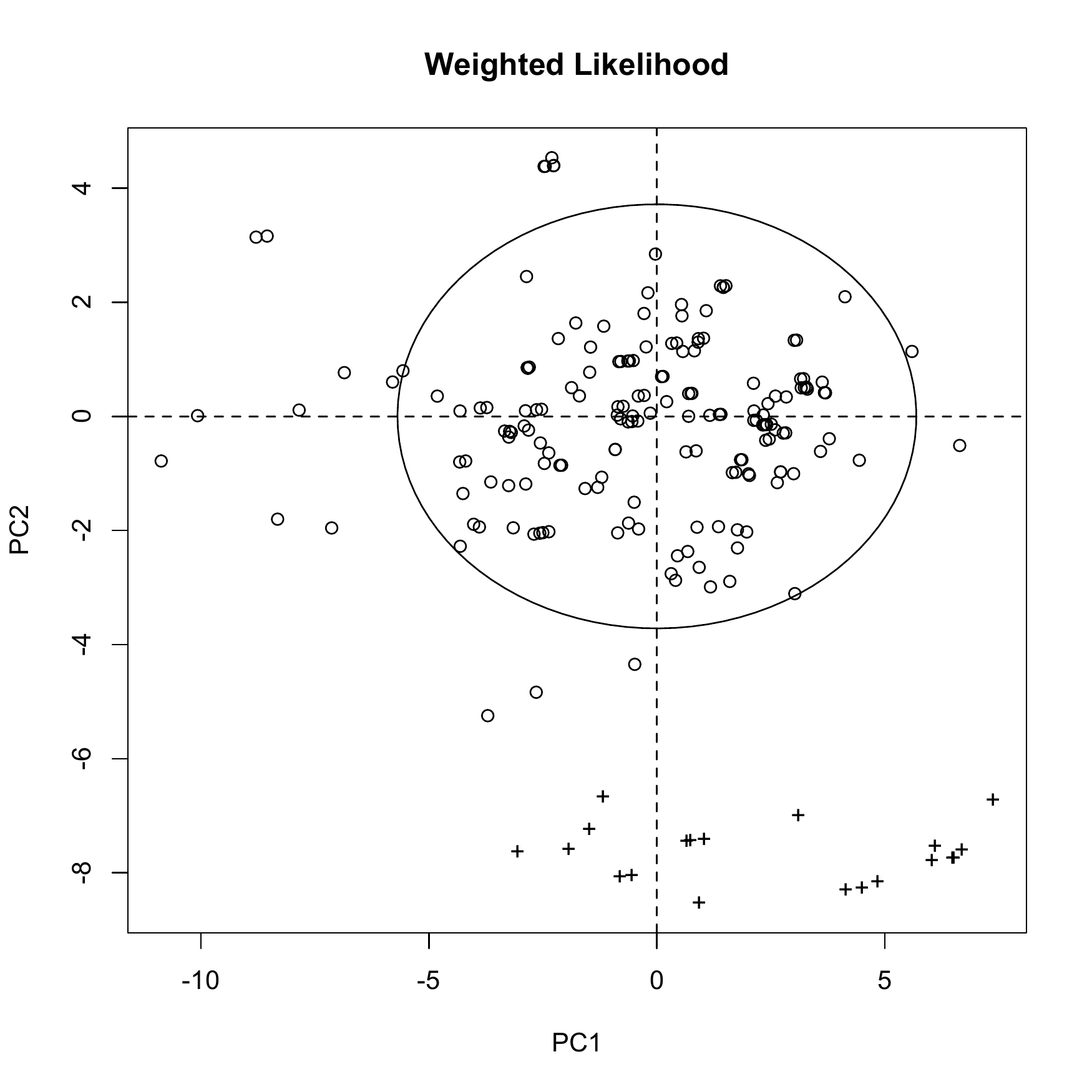}
\includegraphics[width=0.49\textwidth]{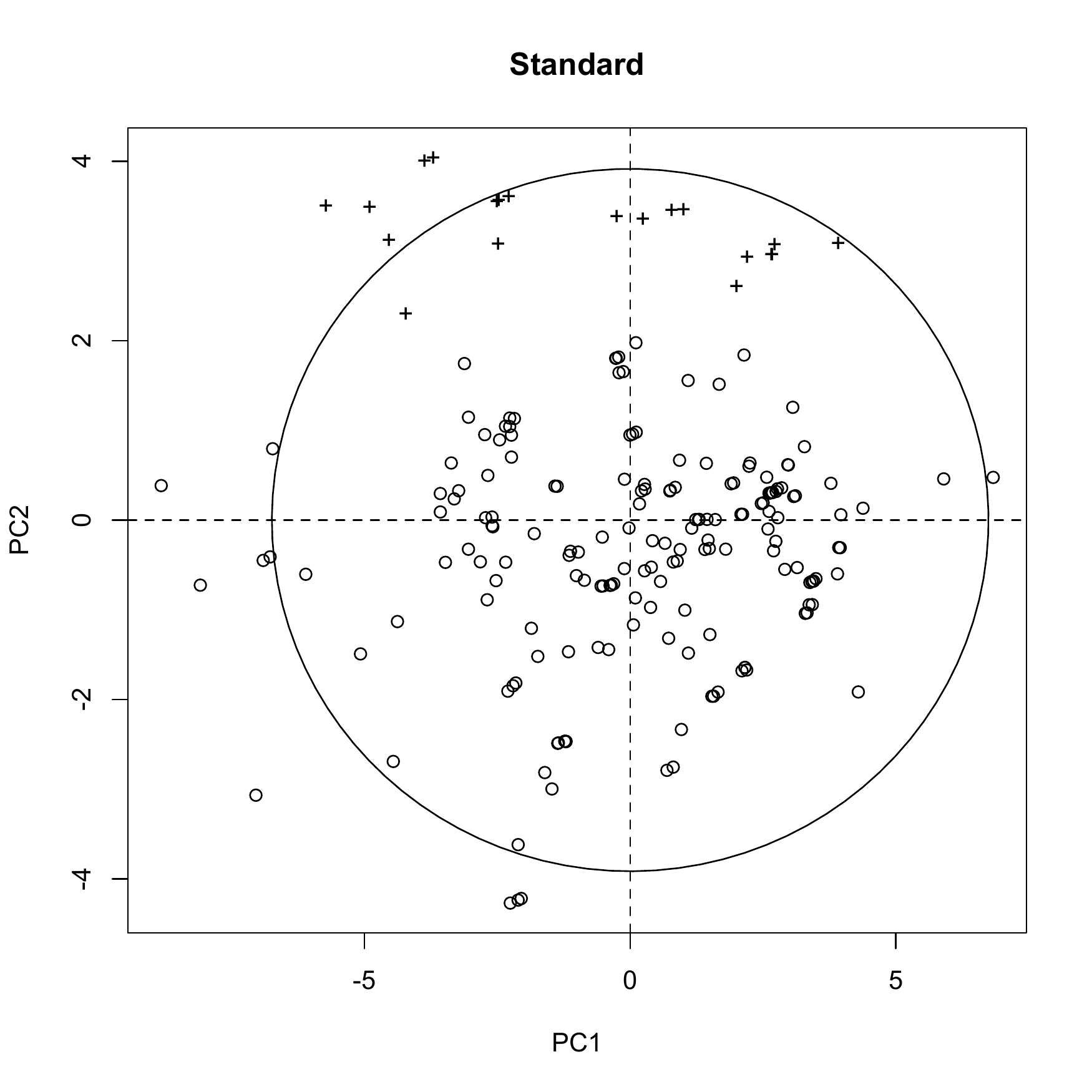}
\includegraphics[width=0.49\textwidth]{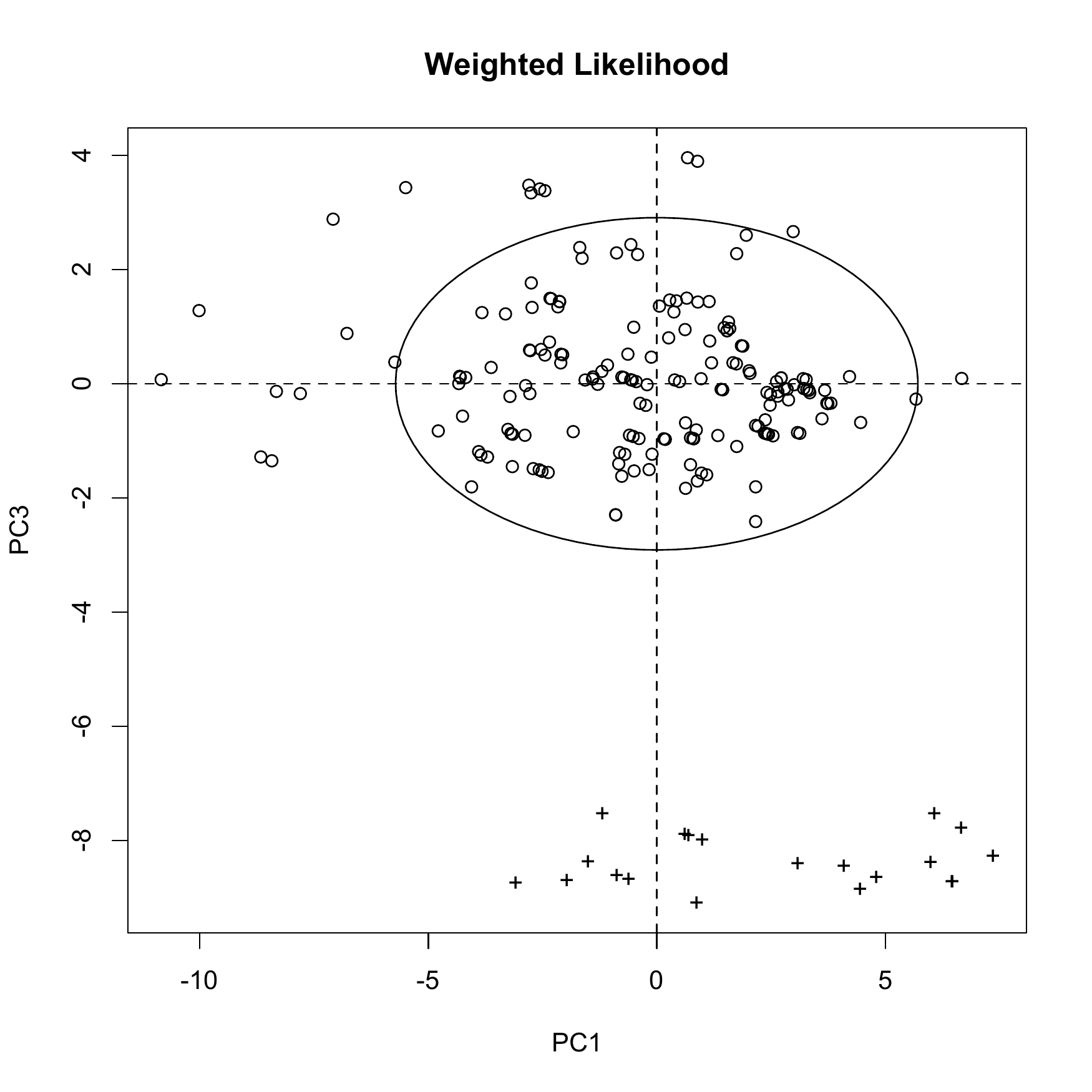}
\includegraphics[width=0.49\textwidth]{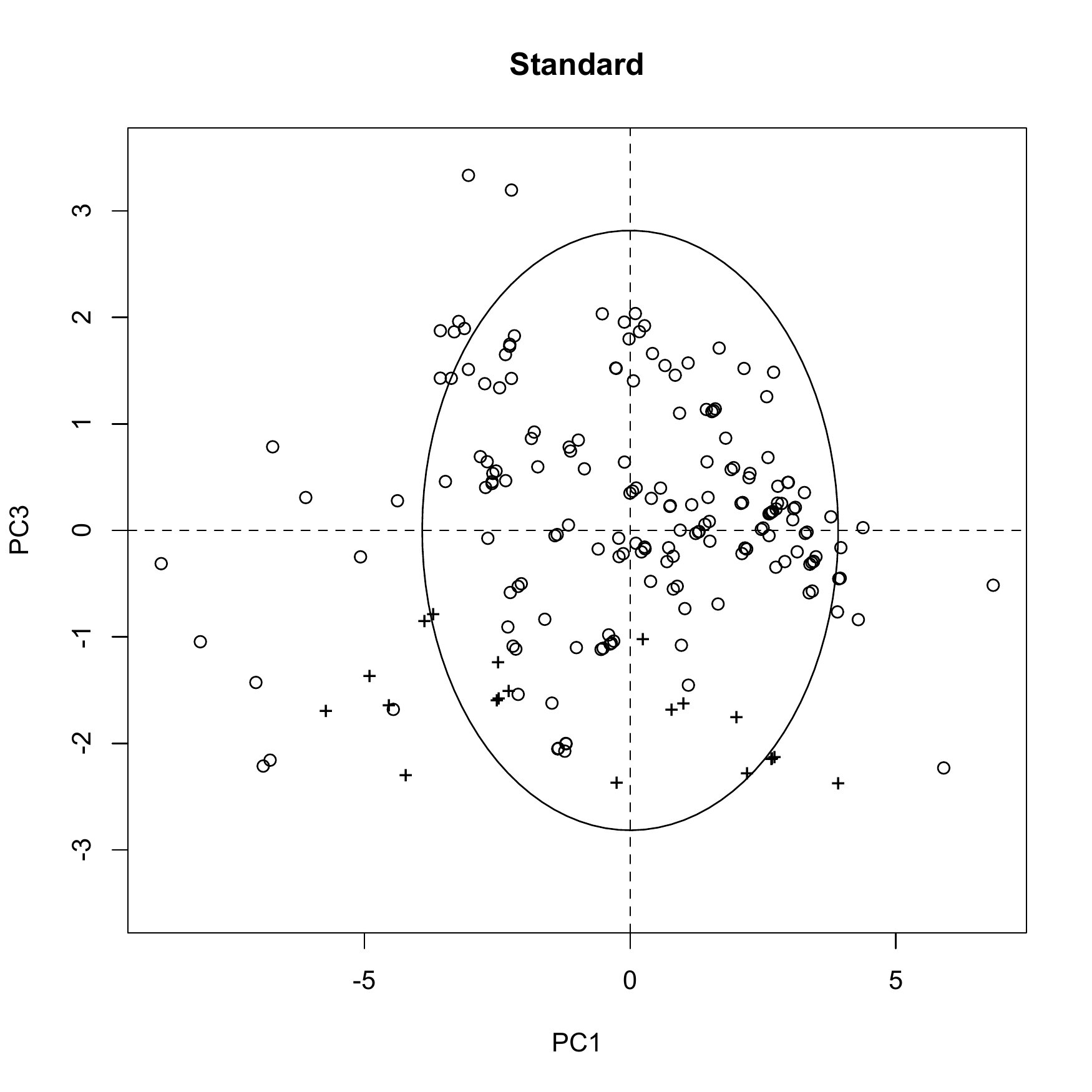}
\includegraphics[width=0.49\textwidth]{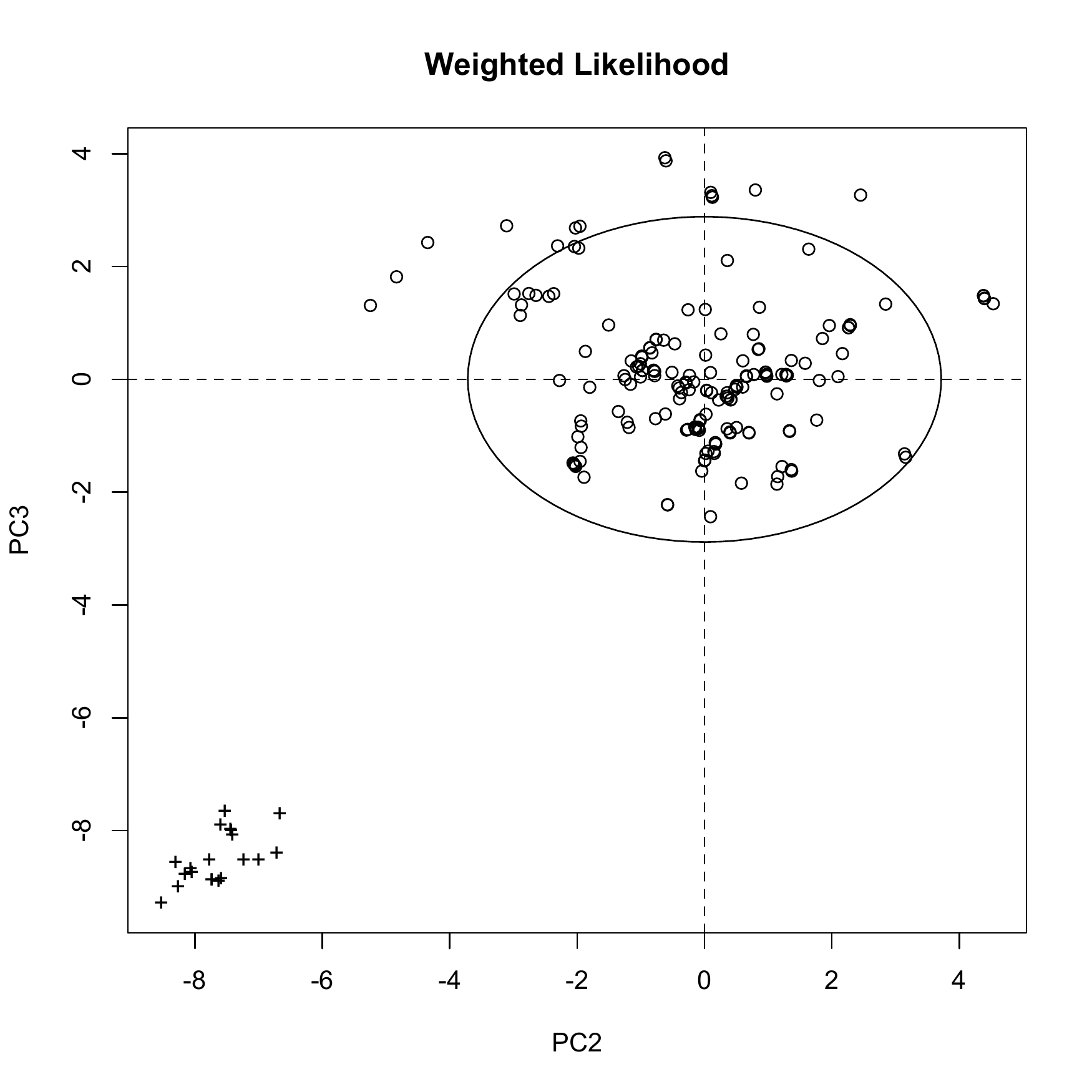}
\includegraphics[width=0.49\textwidth]{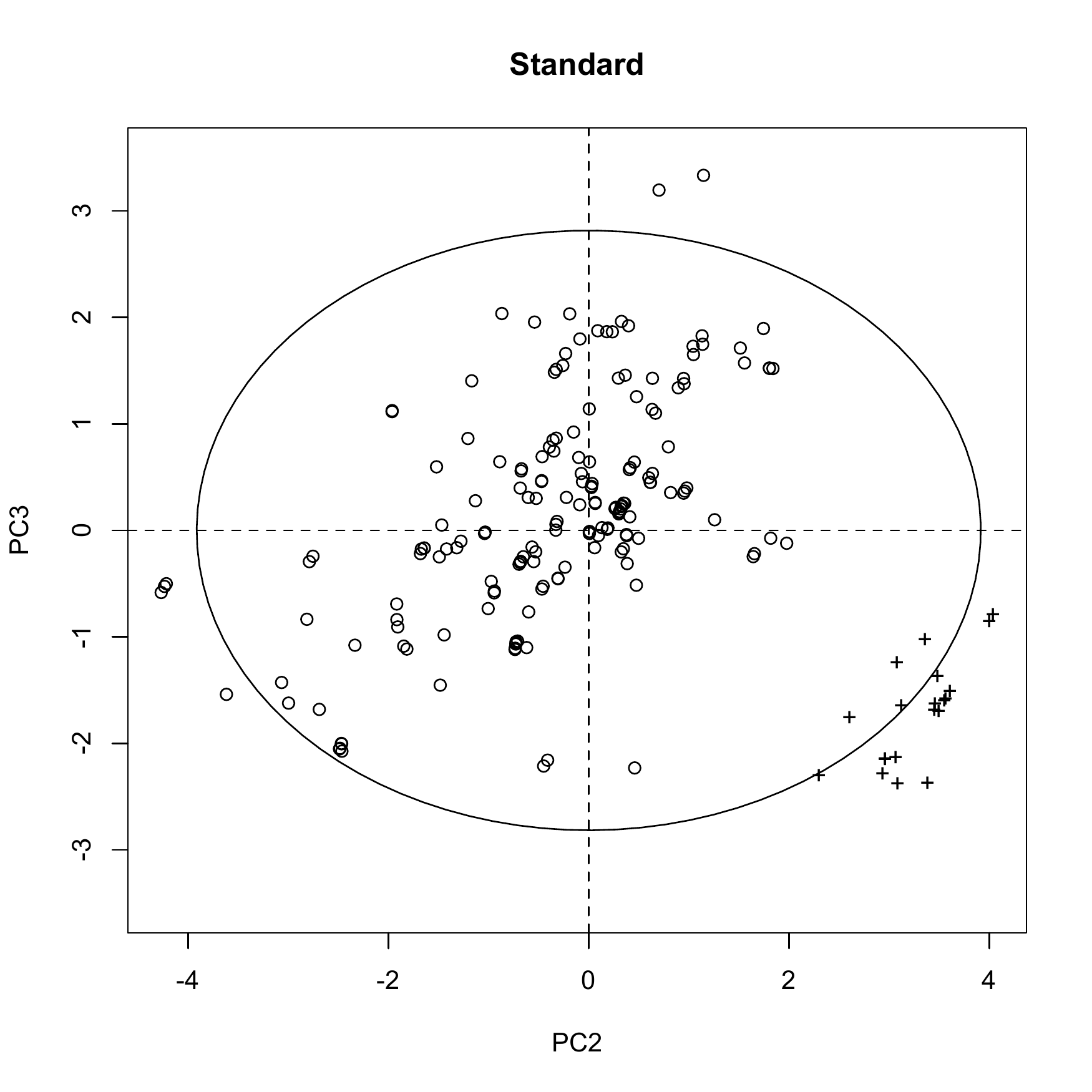}
\caption{Auto data: pairwise score-plots based on the first three components, robust (left) and standard (right). Cars running on a diesel engine are denoted by a $+$.}
\label{auto4}
\end{figure}

Principal Component Analysis (PCA) is undoubtedly the most popular technique for dimension reduction.
The data are projected onto a lower dimensional sub-space so that they are as spread out as possible.
This feature allows to express the covariance structure of the data by means of a small number of new variables (the principal components). These new variables are obtained as linear combination of the original set of variables and are orthogonal each other. The coefficients of the linear combinations are given by the eigenvectors of the covariance matrix and each component accounts for an amount of total variability proportional to the corresponding eigenvalue. PCA is clearly sensitive to the occurrence of outliers, that, in particular, may inflate the variability accounted for by the first components hence leading to wrongly rotated loadings. One approach to robust PCA is based on the eigen-decomposition of a robust estimate of covariance. Here, we employ the WLE to perform a robust PCA on the Auto data. The same example has been discussed in \cite{farcomeni2016robust}. Let us consider the first $k=3$ components. The percentage of explained variance from standard PCA is $76.5\%$ whereas the robust analysis gives a smaller value of $73.6\%$. In order to better explain the deleterious effect of outliers on standard PCA and the effectiveness of our weighted approach,
Figure \ref{auto4} displays the pairwise score-plots based on the first three components. The group of cars running on diesel is clearly spotted by the robust components in the left panels, whereas this does not happens in the right panel. 
A typical effect due to the presence of outliers can be seen in the last panel:
the second and third component from standard PCA still show a linear trend and only the effect of the outlying cars leads to a null correlation. 

\begin{figure}[t]
\centering
\includegraphics[height=0.45\textheight]{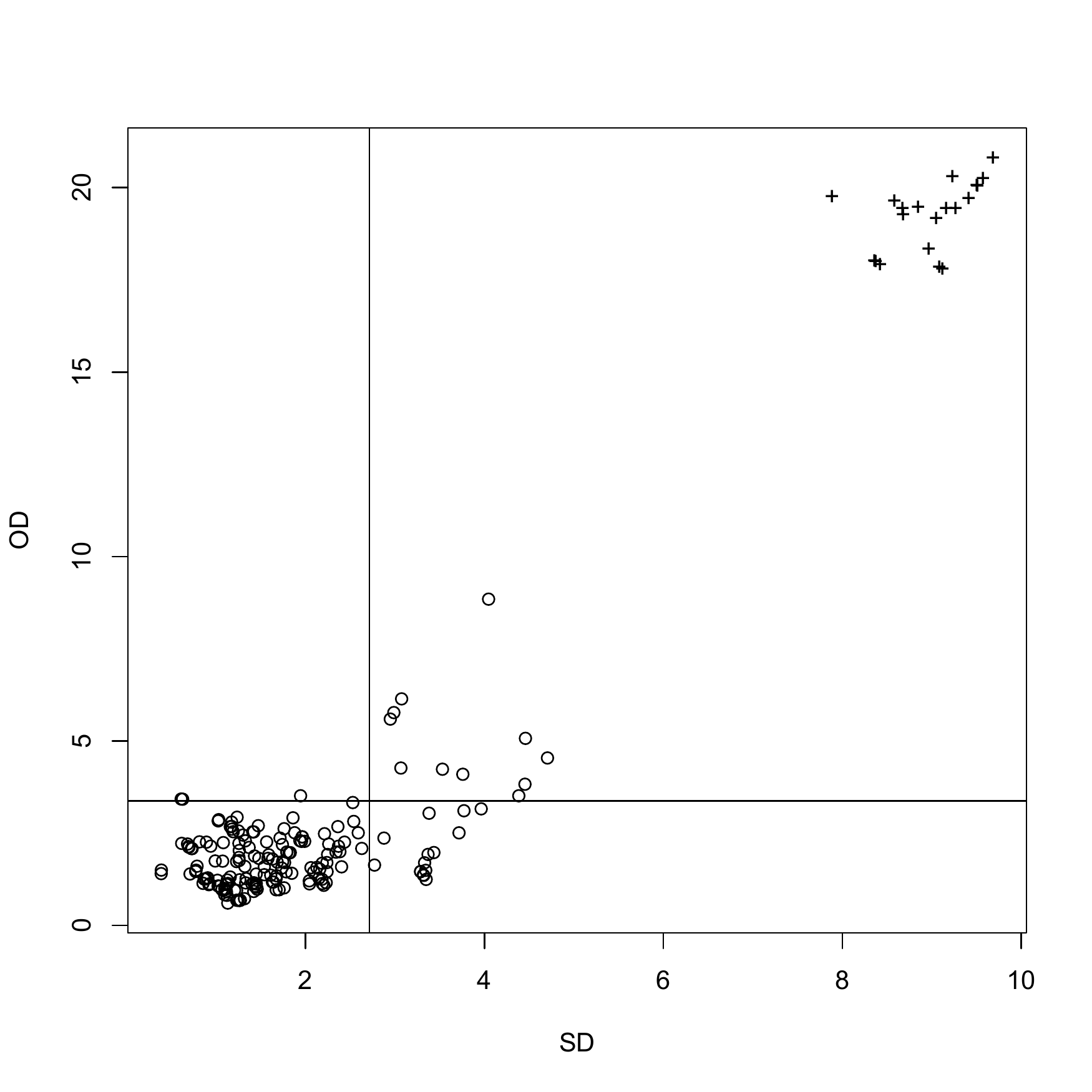}
\caption{Auto data: outlier map based on WLE with $k=3$ components. Cars running on a diesel engine are denoted by a $+$.}
\label{auto5}
\end{figure}

Robust PCA is an effective tool in outlier detection when the dimensionality is not of a manageable size. The usual tool is an outlier map, displayed in Figure \ref{auto5}, that is obtained by plotting for each data point its score and orthogonal distance: the group of outlying cars is clearly separated from the rest but also other 
outlying points are visible.
Guidelines to find the cut-off values are given in \cite{hubert2005robpca}.

We only mention here, that the WLE of multivariate location and scatter could be used in the technique developed by \cite{greco2016plug} to obtain sparse and robust PCA.

\subsection{Discriminant Analysis}
Discriminant analysis is concerned with the problem of assigning data to one of several groups. The observations within each group are assumed to arise from a multivariate normal distribution. In linear discriminant analysis (LDA) it is assumed homogeneity of the covariance matrices, whereas in quadratic discriminant analysis (QDA) the groups are allowed to have different scatters. Let $\pi_j, j=1,2,\ldots, k$ denote the prior probabilities. 
The linear discriminant rule classifies observations by maximizing 
$$
\log\hat\pi_j-\frac{1}{2}d^2(y, \hat\mu_j, \hat\Sigma_p)
$$
and the quadratic discriminant rule classifies observations by maximizing
$$
\log\hat\pi_j-\frac{1}{2}d^2(y, \hat\mu_j, \hat\Sigma_j)
$$
where $\hat\pi_j=\frac{n_j}{\sum_{j=1}^k n_j}$ is an estimate of prior probabilities to be used when prior information is not available, $\hat\mu_j$ is an estimate of the group vector mean, $\hat\Sigma_p$ is a pooled estimate of the common scatter and, $\hat\Sigma_j$ is an estimate of the group scatter matrix. Actually, an appealing approach to define a discriminant function that is not prone to contamination in the data is based on robust estimates of location and common covariance matrix \citep{hubert2004fast, he2000high}. Here, we apply weighted likelihood to perform robust LDA and QDA.
In particular, we consider two different strategies to obtain a robust pooled estimate of the covariance matrix, in a fashion similar to what happens when using the MCD estimator \citep{todorov2007}. By paralleling the standard technique, the first estimate (WLEA) is obtained by averaging the unbiased estimates from each group as follows:
$$
\hat\Sigma_{p}^A=\frac{\sum_{j=1}^k \gamma_j\omega_j\hat\Sigma_{uj}}{\sum_{j=1}^k \gamma_j\omega_j}, \ \omega_j=\sum_{i=1}^{n_j} \hat w_{ij}
$$
The second estimate (WLEB) can be obtained after the following steps. First center the data from each group by a robust estimate of location $\hat\mu_{j0}$; one could use the L1 (spatial) median, for instance. Then, evaluate the WLE $(\hat\mu_p, \hat\Sigma_p^B)$ from all the centered data and update the group vector means as
$\hat\mu_j^B=\hat\mu_{j0}+\hat\mu_p$. The latter approach needs only one robust estimate of covariance rather than one for each group as in the former one. Nevertheless, an alternative, even if slightly more demanding, still consists in running Algorithm \ref{a1} for  each group and centering the data by using the WLE of location from each of them in the first step.

Let us apply weighted LDA and QDA to the Diabetes data. These data consists of three measurement of plasma, {\tt glucose},  {\tt insuline} and {\tt sspg}, made on 145 non-obese adult patients classified into three groups: normal subjects, chemical diabetes and overt diabetes. 
\begin{figure}[t]
\centering
\includegraphics[width=0.32\textwidth]{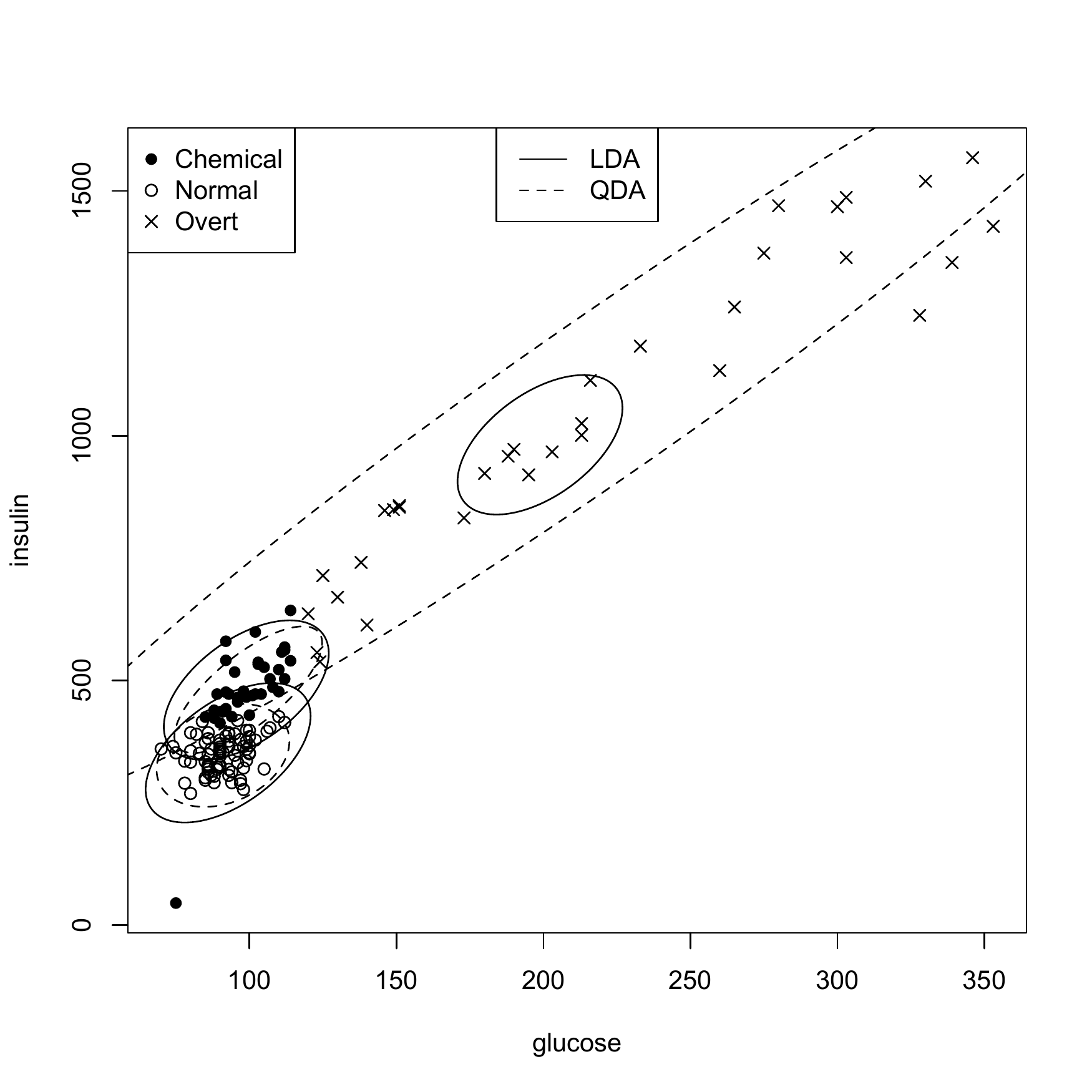}
\includegraphics[width=0.32\textwidth]{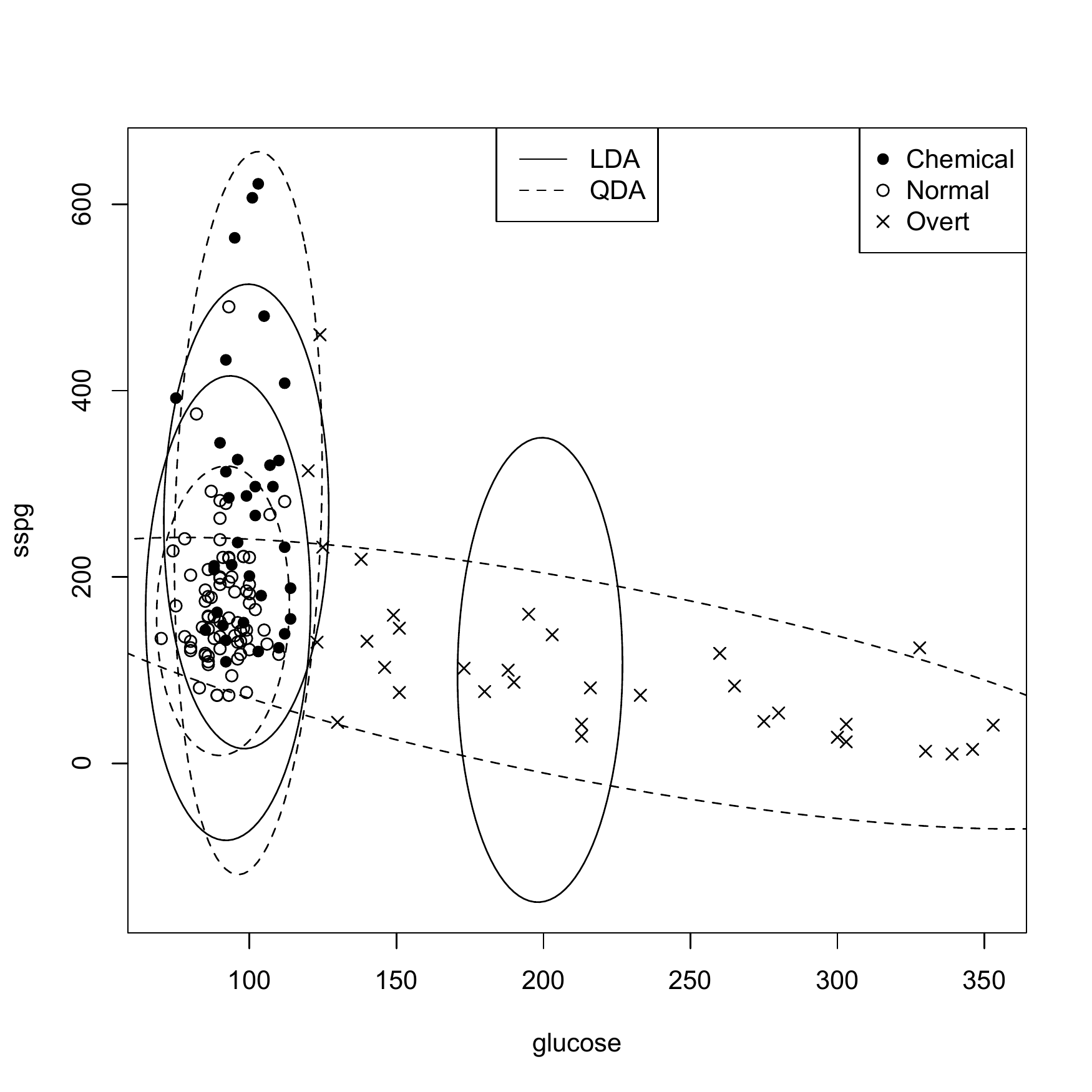}
\includegraphics[width=0.32\textwidth]{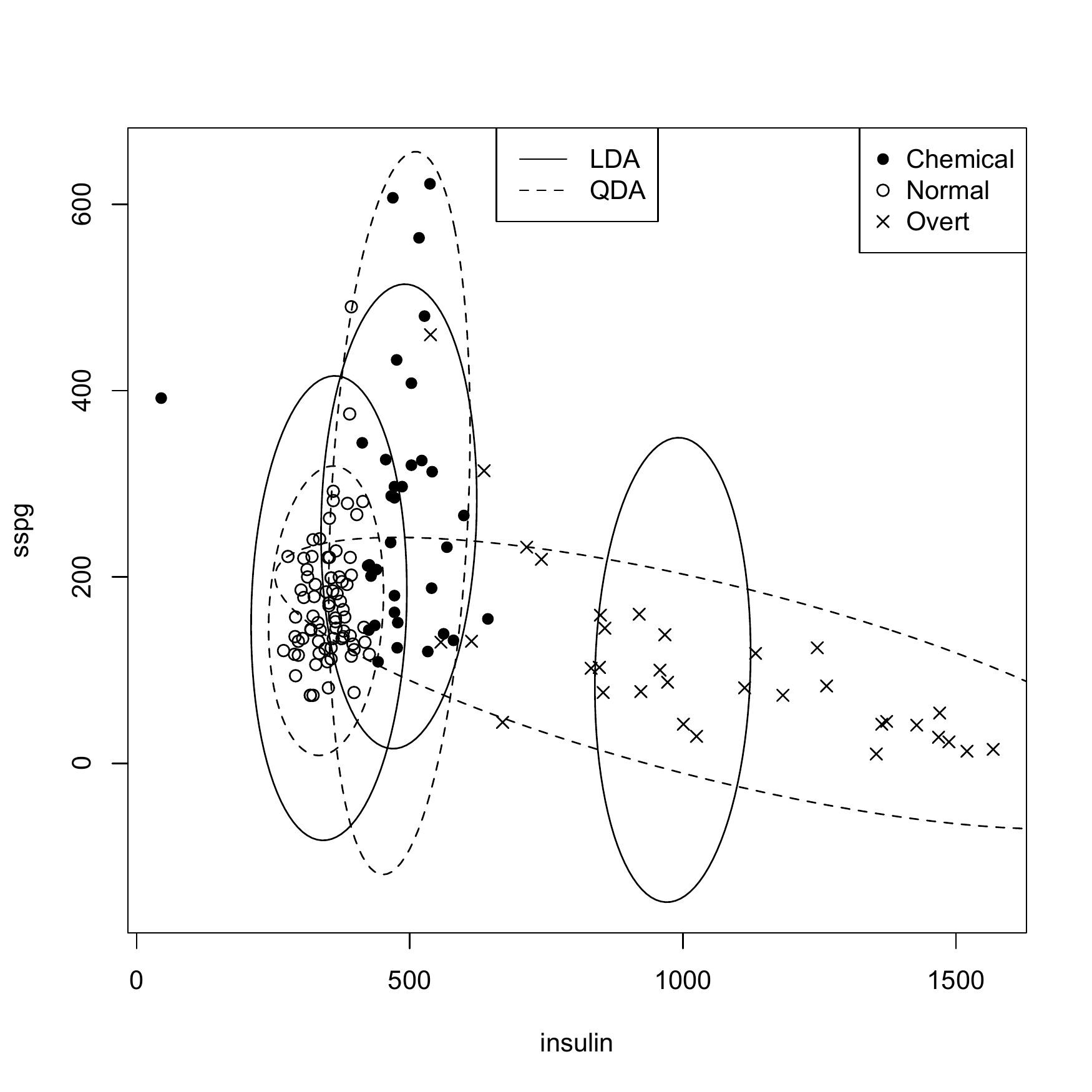}
\caption{Diabetes data: pairwise scatter-plots with 0.975-level tolerance ellipses over-imposed for each group, driven by
LDA based on WLEB and QDA based on group-wise WLEs.}
\label{diab}
\end{figure}  
The data, along with the fitted groups according to LDA based on WLEB and QDA stemming from group-wise WLEs, are displayed in Figure \ref{diab}. The fitted groups appears as 0.975-level tolerance ellipses. It is worth noting the differences among the two techniques concerning, in particular, the peculiar nature of the overt diabetes group.  
Actually, the nature of correlation between {\tt glucose} and {\tt sspg} and {\tt insuline} and {\tt sspg} in the third group is different from what happens in the other two groups.
The entries in Table \ref{tabDA} give the estimates of the misclassification rate based on all the data (ALL) and on leave-one-out cross validation (CV) based on the WLE, MLE and MCD, for LDA and QDA. The use of robust estimates of multivariate location and scatter improves classification accuracy over the standard approach based on the MLE and the WLE performs satisfactory compared to the MCD. In particular, both LDA based on WLEB and QDA stemming from the group-wise WLEs lead to the same results.

\begin{table}[t]
\caption{Diabetes data: misclassification rates for different rules based on all the data and on leave-one-out cross validation.}
\label{tabDA}
\centering
\begin{tabular}{crcc}
&&ALL & CV\\
\hline
&MLE&0.131&0.131\\
&WLEA&0.124&0.131\\
LDA&WLEB&0.076&0.103\\
&MCDA&0.124&0.131\\
&MCDB&0.083&0.117\\
\hline
&MLE&0.076&0.110\\
QDA&WLE&0.076&0.103\\
&MCD&0.083&0.103\\
\hline
\end{tabular}
\end{table}

\section*{Appendix}
{\bf Lemma 1}
Let $A=\omega\hat\Sigma=\omega\gamma\hat\Sigma_u$, $A_{(-i)}=\omega_{(-i)}\hat\Sigma_{(-i)}=\omega_{(-i)}\gamma_{(-i)}\hat\Sigma_{u(-i)}$, with $\omega=\sum_{i=1}^n w_i$ and the subscript $(-i)$ denote that the $i$-th observation has been removed. Then
\begin{equation}
\label{lemma1}
A-A_{(-1)}=\frac{\omega}{\omega_{(-i)}}(Y_i-\hat\mu)(Y_i-\hat\mu)^{\T}w_i \ .
\end{equation}

{\it Proof}. It is known that:
\begin{eqnarray*}
A&=&\sum_{j\neq i} Y_jY_j^{\T}w_i+ Y_iY_i^{\T}w_i-\omega\hat\mu\hat\mu^{\T} \\
A_{(-i)}&=&\sum_{j\neq i} Y_jY_j^{\T}w_i-\omega_{(-i)}\hat\mu_{(-i)}\hat\mu_{(-i)}^{\T} \\
\end{eqnarray*}
Then
$$
A-A_{(-i)}=Y_iY_i^{\T}w_i-\omega\hat\mu\hat\mu^{\T}+\omega_{(-i)}\hat\mu_{(-i)}\hat\mu_{(-i)}^{\T},
$$
and by replacing
$$
\omega\hat\mu=\omega_{(-i)}\hat\mu_{(-i)}+Y_iw_i
$$
the equality stated in (\ref{lemma1}) is obtained.
\\

{\it Proof of Proposition 1}.
Let us consider the random quantity
\begin{equation*}
\label{Ratio}
R_{(-i)}=\frac{|A_{(-i)}|}{|A|} \ .
\end{equation*}

By using Lemma 1, we find that
\begin{eqnarray*}
\label{p1}
R_{(-i)}&=&\left[ 1+ \frac{\omega}{\omega_{(-i)}}(Y_i-\hat\mu)^{\T}A_{(-1)}^{-1}(Y_i-\hat\mu)w_i\right]^{-1} \nonumber\\ 
&=&\left[ 1+ \frac{\omega}{\omega^2_{(-i)}\gamma_{(-i)}}(Y_i-\hat\mu)^{\T}\hat\Sigma_{u(-1)}^{-1}(Y_i-\hat\mu)w_i\right]^{-1} \nonumber \\ 
&=&\left[ 1+ X_i^{\T}\hat\Sigma_{u(-1)}^{-1}X_i\right]^{-1} 
\end{eqnarray*}
where $X_i=\sqrt{\frac{\omega w_i}{\omega^2_{(-i)}\gamma_{(-i)}}}(Y_i-\hat\mu)$.
At the assumed multivariate normal model, by applying the asymptotic results on the distribution of the WLE and the behavior of the weights, we have that  
$$
(Y_i-\hat\mu)\stackrel{d}{\rightarrow}N_p\left(0, \frac{n-1}{n}\Sigma\right)
$$
and $$\sqrt{\frac{\omega w_i}{\omega^2_{(-i)}\gamma_{(-i)}}}\stackrel{p}{\rightarrow}\sqrt{\frac{n}{(n-1)(n-2)}}.$$ Then, from the Slutsky theorem,
$X_i\stackrel{d}{\rightarrow}\frac{1}{\sqrt{n-2}}N(0,\Sigma)$. 

In a similar fashion, at the assumed multivariate normal model, we can state that $\hat\Sigma_{u(-1)}\stackrel{d}{\rightarrow}W_p(\Sigma, n-1)$. 
By using the fact that, conditioning on the weights, $X_i$ and $\hat\Sigma_{u(-i)}$ are independent, since $\hat\Sigma_{u(-i)}$ does not involve $Y_i$, we have that, asymptotically, the distribution of $X_i^{\T}\hat\Sigma_{u(-i)}^{-1}X_i$ is the Hotelling distribution.
Then, by using standard results, 
it is possible to derive the asymptotic distribution of  $R_{(-i)}$, that is
$$
R_{(-i)}\stackrel{d}{\rightarrow} \textrm{Beta}\left(\frac{n-p-1}{2}, \frac{p}{2}\right) \ .
$$
According to Lemma 1, we are also allowed to write
\begin{eqnarray*}
R_{(-i)}&=&1-\frac{w_i}{\omega_{(-i)}\gamma}(Y_i-\hat\mu)^{\T}\hat\Sigma_u^{-1}(Y_i-\hat\mu)\\
&=&1-\frac{w_i}{\omega_{(-i)}\gamma}d^2(Y_i,\hat\mu,\hat\Sigma_u)
\end{eqnarray*}
and
\begin{equation*}
\label{final}
d^2(Y_i,\hat\mu,\hat\Sigma)=\left(1-R_{(-i)}\right)\frac{\omega_{(-i)}\gamma}{w_i} \ .
\end{equation*}
Since $(1-R_{(-i)})\stackrel{d}{\rightarrow} \textrm{Beta}\left(\frac{p}{2},\frac{n-p-1}{2}\right) $
and $\frac{\omega_{(-i)}\gamma}{w_i}\stackrel{p}{\rightarrow}\frac{(n-1)^2}{n}$, then, again by Slutsky theorem, the result stated in (\ref{prop1}) is obtained and the proof is concluded.


\end{document}